\documentclass[aps,preprint,epsfig,rotate]{revtex4}

\usepackage{graphicx}
\usepackage{bm}
\usepackage{epsfig}

\input epsf
\begin{document}
\title{Bound state properties and photodetachment of the negatively charged hydrogen ions.}

 \author{Alexei M. Frolov}
 \email[E--mail address: ]{afrolov@uwo.ca}

\affiliation{Department of Applied Mathematics \\
 University of Western Ontario, London, Ontario N6H 5B7, Canada}

\date{\today}

\begin{abstract}

Absorption of infrared and visible radiation from stellar emission spectra by the negatively charged hydrogen ions 
H$^{-}$ is considered. The explicit formula for the photodetachment cross-section of the negatively charged hydrogen 
ion(s) is derived. Photodetachemnt cross-sections of the ${}^{\infty}$H$^{-}$, ${}^{3}$H$^{-}$ (or T$^{-}$), 
${}^{2}$H$^{-}$ (or D$^{-}$) and ${}^{1}$H$^{-}$ ions are determined to high accuracy and for a large number of 
photo-electron momenta/energies. We introduce criteria which can be used to evaluate the overall quality of highly 
accurate wave functions of the hydrogen ion(s). One of these criteria is based on highly accurate calculations of the 
lowest order QED corrections in the negatively charged hydrogen ions, including ${}^{1}$H$^{-}$ (protium), ${}^{2}$H$^{-}$ 
(deuterium), ${}^{3}$H$^{-}$ (tritium) and model ion with the infinitely heavy nucleus ${}^{\infty}$H$^{-}$. An effective 
approach has been developed to calculate three-body integrals with the Bessel functions of different orders. Some 
preliminary evaluations of the phototdetachment cross-sections of the negatively charged hydrogen ions are performed. 
Inverse bremsstrahlung in the field of the neutral hydrogen atom is briefly discussed.



\end{abstract}

\maketitle
\newpage

\section{Introduction}

As is well known (see, e.g., \cite{Sob} - \cite{Zir}) the negatively charged hydrogen ions ${}^{1}$H$^{-}$ (protium) and ${}^{2}$H$^{-}$ (deuterium 
or D$^{-}$) are of great interest in Stellar Astrophysics. Indeed, the negatively charged hydrogen ions substantially determine the absorption of
infrared and visible radiation in photospheres of all stars with temperatures bounded between $T_{max} \approx$ 8,250 $K$ (late A-stars) and $T_{min} 
\approx$ 2,750 $K$ (early M-stars). The main contribution to the light absorption by the negatively charged hydrogen ion H$^{-}$ comes from its 
photodetachment
\begin{equation}
 {\rm H}^{-} + h \nu = {\rm H} + e^{-} \label{e0}
\end{equation}
where the notation $h \nu$ designates the incident light quantum, while $e^{-}$ means the free electron, or photo-electron, for short. The notation $h$ 
stands for the Planck constant. Note that each of the negatively charged hydrogen ions, i.e. the ${}^{1}$H$^{-}$, ${}^{2}$H$^{-}$ (or D$^{-}$), 
${}^{3}$H$^{-}$ (or T$^{-}$) and ${}^{\infty}$H$^{-}$ has only one bound (ground) $1^1S-$state with $L = 0$, where $L$ is the angular 
momentum of this two-electron ion (see, e.g., \cite{Fro2005}). The ground $1^1S-$state is a singlet state, i.e., the total electron spin $S$ in this 
state equals zero, and therefore, $2 S + 1 = 1$. Here and everywhere below in this study the notation ${}^{\infty}$H/${}^{\infty}$H$^-$ stands for the 
model atom/ion with the infinitely heavy nucleus. The total non-relativistic energy of the ground $1^1S-$state in the ${}^{\infty}$H$^-$ ion is -0.527751 
016544 377196 59055(5) $a.u.$ \cite{Fro2005}, while the total energy of the hydrogen atom with the infinitely heavy nucleus (${}^{\infty}$H) is -0.5 $a.u.$ 
Therefore, the binding energy of the ${}^{\infty}$H$^-$ ion is $\chi_1 \approx$ 0.027 751 $a.u.$ $\approx$ 0.755 143 9 $eV$ and the corresponding ionization 
frequency $\nu_1 = \frac{\chi_1}{h}$ is located in the far infrared part of the spectrum. Note that the energies of `visible' light quanta are $\approx$ 2 - 
3 $eV$, while the total energy of the ground state of hydrogen atom is $\approx$ 13.605 $eV$. Light quanta with such energies ($\approx$ 13.605 $eV$) 
correspond to the vacuum ultraviolet region. 

The final atomic state which arises after photodetachment of the negatively charged hydrogen ion, Eq.(\ref{e0}), includes the neutral hydrogen atom H and `free' 
electron $e^{-}$. The `free' electron moves in the field of the central hydrogen atom. The kinetic energy of this electron depends upon the total energy of the 
incident photon and can, in principle, be arbitrary. Formally, all radiation quanta with the frequencies $\nu \ge \nu_1 = \frac{\chi_1}{h}$ can produce
photodetachment of the negatively charged hydrogen ion H$^{-}$. However, it can be shown (e.g., see below) that the photodetachment cross-section rapidly 
(exponentially) vanishes, if the photon energy substantially exceeds numerical values of the ionization potential of the H$^{-}$ ion. Another reason which allows 
one to ignore the role of high-energy photons in the H$^{-}$ ion photodetachment follows from stellar Astrophyscs. Indeed, if we assume that  the actual stellar 
photosphere is at thermal equilibrium with $T \approx$ 4000 - 8000 $K$, then the part of `high-energy' photons which can produce excitations of the final hydrogen 
atom is extremely small. For instance, to produce photodetachment of the H$^{-}$ ion with the formation of the final H atom in the $2s-$state the `photon 
temperature' $T_f$ must be $\ge$ 78944 $K$. Analogous `photon temperature' to form the final H atom in the $3s-$state is around 118,416 $K$. For our Sun the 
fraction of the `hot' thermal photons with $T \ge$ 78944 $K$ is very small ($\le 1 \cdot 10^{-3}$ \%). Even in photospheres of late $A-$stars the fraction of `hot' 
thermal photons is less than 0.5 \%. In reality, such photons become noticeable in photospheres of B-stars with surface temperature $T \ge 20,000 K$. However, in 
classical B-stars the photodetachment of the H$^{-}$ ions does not play any role and most of the emitted radiation is absorbed by neutral hydrogen and helium 
atoms. This explains why the photodetachment of the H$^{-}$ ion in our Sun and in the late A-stars and in all F, G and K stars and in some other stars always proceeds 
with the formation of the final H atom in its ground (bound) $1s-$state. In the hot O- and B-stars, with surface temperatures $T_s \ge 28,500$ $K$ and $T_s \ge 11,000$ 
$K$, respectively, the absorption of radiation quanta is mainly related to the transitions between different bound states in helium and hydrogen atoms. The 
negatively charged H$^{-}$ ion does not exist at such temperatures as a stable (or bound) system.

This situation can be different in the Be-stars, i.e. in the B-stars with the rapidly rotating, low-dense disk of hydrogen atoms \cite{Sob}, \cite{Malb}. Illumination 
of this disk by radiation emitted by the central hot star produces excitation of hydrogen atoms in the stellar disk \cite{Cass}. Excitation of hydrogen atoms and 
following emission of absorbed radiation generates very special optical spectra of the Be-stars, which contain thermal (continous) spectra from the central star (or 
central cluster) and linear spectra which corresponds to the transition between excited states in the hydrogen atoms from the disk. For such stars photodetachment of 
the H$^{-}$ ions with the formation of the final hydrogen atoms in their excited states may be important in the outer areas of stellar disk. However, for all Be-stars
photodetachment of the H$^{-}$ ions may proceed in the spatial areas located at the large distances ($\ge$ 150,000,00 $km$) from the hot central star, i.e. in areas 
which are far away from stellar photospheres. Below in this study we shall not discuss photodetachment of the H$^{-}$ ion at such `exotic' conditions.

On the other hand, let us note that the H$^{-}$ ions play practically no role for the cold M and N stars with $T_s \le$ 2,700 $K$ where the absorption of radiation by
various molecular species and atoms of metals becomes important. However, in the late F, G and early K stars the absorption of infrared and visible radiation by the 
negatively charged hydrogen ions is maximal. This includes our Sun which is a star of spectral type G2. The great role of the H$^{-}$ ions for our Sun was suggested by R. 
Wild in 1939 (see discussions and references in \cite{Sob}, \cite{Aller} and \cite{Zir}). An effective absorbtion of large amount of infrared solar radiation by the 
negatively charged hydrogen ions is crucial to reach a correct thermal balance at our planet. 

In general, the light absorption by an arbitrary atomic system is determined by separate contributions from the corresponding bound-bound, bound-free and free-free 
transitions in this system. As mentioned above the negatively charged hydrogen ion has only one bound (ground) $1^1S-$state, i.e.  for the H$^{-}$ ion there is no 
contribution from the bound-bound transitions. The bound-free transitions in the H$^{-}$ ion correspond to the photodetachment, Eq.(\ref{e0}), while the free-free 
(electron) transitions formally represent the radiation-stimulated electron scattering in the field of the neutral hydrogen atom (or inverse bremsstrahlung, for short)
\begin{equation}
 e^{-} + h \nu + {\rm H} = {\rm H} + e^{-} \label{ee0}
\end{equation}
where the kinetic energy of the final electron differs from its incident energy due to the photon's absorption. If the kinetic energy of the both incident photon and 
electron in Eq.(\ref{ee0}) are small, then the final state of the hydrogen atom coincides with its original (ground) state. For our Sun the role of inverse 
bremsstrahlung is substantial for infra-red photons with $\lambda \approx$ 12,000 - 20,000 $\AA$.  

In this study we determine the photodetachment cross-section of the negatively charged hydrogen ion (or H$^{-}$ ion). In particular, we want to increse substantially the 
overall accuracy of numerical calculations of the photodetachment cross-section of the H$^{-}$ ion. Currently, the photodetachment cross-section of the H$^{-}$ ion must 
be determined to very high accuracy ($\pm$ 0.1 - 0.2 \%). The moderate accuracy known from earlier works ($\pm$ 2 \% - 10 \%) is not sufficient to describe effects directly 
related to the climate change and surface temperature variations at our planet. In our calculations we apply our highly accurate bound state wave functions constructed for 
the two-electron H$^{-}$ ion. The crucial part of the method is analytical and numerical computations of three-particle integrals which contain spherical Bessel functions 
of the first kind. Here we solved this problem and now we can determine a large number of different three-particle integrals with Bessel functions $j_L(p_e r)$ to very high 
accuracy. This paper opens a new avenue in accurate numerical computations of the photodetachment cross-sections of negatively charged hydrogen ions: ${}^{\infty}$H$^{-}$, 
${}^{3}$H$^{-}$ (or T$^{-}$), ${}^{2}$H$^{-}$ (or D$^{-}$) and ${}^{1}$H$^{-}$. By using our current methods we can determine the photodetachment cross-sections for all
these ions of hydrogen isotopes. Applications of our procedure are not limited to the cases of the ground state of the hydrogen atom(s), but also include a number of  
excited states. We also discuss inverse bremsstrahlung, i.e. radiation emitted by an electron which moves in the field of the hydrogen atom and a large number of bound state 
properties of the negatively charged hydrogen ions. The computed properies allows us to determine the lowest-order QED corrections for these ions.  

\section{Photodetachment of the H$^{-}$ ion}

In this Section we derive and discuss the explicit formulas which are used for numerical computation of the photodetachment cross-section of the H$^{-}$ ion. Note that the 
absorption of radiation by the negatively charged hydrogen ions has been investigated in numerous earlier studies (see, e.g., \cite{Chand} - \cite{FroSm03} and references 
therein). An extensive review of different experimental and computational procedures used to study the interaction of light with the hydrogen ions can be found in \cite{PRA2014}.
This paper \cite{PRA2014} also contains all essential references related to this problem up to 2014. In general, the photodetachment corresponds to the bound-free (optical) 
transitions, since the final state of one of the two electrons is an unbound state, or, in other words, the state from the unbound spectra of the H$^{-}$ ion. For the two-electron 
H$^{-}$ ion it is possible to obtain the closed analytical formula for the photodetachment cross-section, if the non-relativistic (or dipole) approximation is used. The derivation 
of this formula is considered below in this Section. 

Let us assume that the incident (i.e. second) electron was bound to the neutral hydrogen atom with the binding energy $\epsilon_{i} = -I$, where $I = \chi_1$ is the ionization 
potential of the H$^{-}$ ion. The incident photon has the momentum ${\bf k}$ and the frequency $\omega$ (or energy $\hbar \omega = \omega$). In the final state the emitted 
photo-electron moves as a free particle with the momentum ${\bf p}$ and energy $\epsilon_f$. Since ${\bf p}$ is a continuous variable, the photodetachment cross-section is written 
in the form \cite{AB}
\begin{equation}
 d\sigma = 2 \pi \mid {\cal A}_{i \rightarrow f} \mid^2 \delta(-I + \omega - \epsilon_f) \frac{d^3 p}{(2 \pi)^3} \label{cross}
\end{equation}
where ${\cal A}_{i \rightarrow f}$ is the transition amplitude (see, Eq.(\ref{Matrix}), below) and the wave function 
of the final state is normalized per one particle in unit volume $V = 1$. All wave functions and expressions used in computations of the transition amplitude 
${\cal A}_{i \rightarrow i}$ are assumed to have a proper permutation symmetry upon spin-spatial coordinates of the two electrons 1 and 2. This allows one to 
write the formula, Eq.(\ref{cross}), and all formulas below in one-electron form.

The delta-function in Eq.(\ref{cross}) is excluded by integrating over the momenta ${\bf p}$ of the photo-electron. Indeed, by using 
the formula $d^3{\bf p} = p^2 d\mid {\bf p} \mid do = \epsilon \mid {\bf p} \mid d\epsilon do$ and performing the integration over 
$d\epsilon$ we can remove this delta-function and obtain the following general expression for the cross-section
\begin{equation}
 d\sigma = \frac{\epsilon \mid {\bf p} \mid}{(2 \pi)^2} \mid {\cal A}_{i \rightarrow f} \mid^2 do \label{dcross}
\end{equation}
In the non-relativistic approximation ($\hbar \omega \ll m_e c^2 = E_e \approx \epsilon$) used below we can replace $\epsilon$ by $m_e$. The derivation of the analytical formula for 
the transition amplitude ${\cal A}_{i \rightarrow f}$ is drastically simplified (see, e.g., \cite{AB}) with the use of transverse gauge for the vector potential ${\bf A}$, i.e. 
$div{\bf A} = 0$. In this case the transition amplitude ${\cal A}_{i \rightarrow f}$ is written as a scalar product of the vector potential ${\bf A}$ and transition current 
${\bf j}_{i \rightarrow f}$, i.e. ${\cal A}_{i \rightarrow f} = - e {\bf A} \cdot {\bf j}_{i \rightarrow f}$. The expressions for the vector potential ${\bf A}$ (in the case when 
$div {\bf A} = 0$) and for the transition current ${\bf j}_{i \rightarrow f}$ can be found in standard QED books (see, e.g., \cite{AB}). With these expressions we can write for the 
transition amplitude 
\begin{eqnarray}
 {\cal A}_{i \rightarrow f} = - e {\bf A} \cdot {\bf j}_{i \rightarrow f} = - e \sqrt{\frac{2 \pi}{\omega}} M_{i \rightarrow f} \; \; \; , \; \; 
\end{eqnarray}
where 
\begin{eqnarray}
 M_{i \rightarrow f} = \int \int \psi^{*}_f(1,2) ({\bf e} \cdot {\bf \alpha}_1) \exp(\imath {\bf k}_{f} \cdot {\bf r}_1) \psi_i(1,2) d^3{\bf r}_1 
 d^3{\bf r}_2 \label{Matrix}
\end{eqnarray}
and $\psi_i(1,2)$ and $\psi_f(1,2)$ are the wave functions of the incident and final atomic systems. These two functions are assumed 
to be properly symmetrized in respect to all possible electron-electron permutations and each of these functions has a unit norm. Also, 
in this equation $\alpha_1$ are the three Dirac matrixes of the first electron, ${\bf e}$ and ${\bf k}_{f}$ are two vectors which describe 
polarization and direction of propagation of the incident photon, respectively.  

In this study all electrons are considered as non-relativistic particles, while the energy of the incident photon $\hbar \omega (= \omega)$ 
is assumed to be larger than $I$, but substantially smaller than the energy of electron at rest, i.e. $\hbar \omega \ll m_e c^2$. This 
means that the final velocity of the photoelectron is small, i.e. we are dealing with the non-relativistic problem. Therefore, the Dirac's 
$\alpha$ matrixes in Eq.(\ref{Matrix}) can be replaced by the corresponding components of the velocity ${\bf v}_1 = -\frac{\imath}{m_e} 
\nabla_1$. Moreover, in the non-relativistic approximation we can replace the factor $\exp(\imath {\bf k}_f \cdot {\bf r}_1)$ by uinty. This 
gives the following formula for the photodetachment cross-section
\begin{eqnarray}
 d\sigma_{\nu} = \frac{e^2 \mid {\bf p}_e \mid}{2 m_e \pi \omega} \Bigl| \Bigl\{\int \int 
 \Psi^{*}_{fi}({\bf r}_1, {\bf r}_2) \Bigl[ {\bf e} \cdot \Bigl(\frac{\partial}{\partial {\bf r}_1} + \frac{\partial}{\partial 
 {\bf r}_2}\Bigr) \Bigr] \Psi_{{\rm H^{-}}}({\bf r}_1, {\bf r}_2) d^3{\bf r}_1 d^3{\bf r}_2 \Bigr\} \Bigr|^2 do \; \; \label{knu}
\end{eqnarray}
where ${\bf e}_f$ is the vector which represents the polarization of the incident photon and ${\bf p}_e$ is the momentum of the photo-electron. In 
the case of photodetachment, Eq.(\ref{e0}), the notation  $\Psi_{fi}({\bf r}_1, {\bf r}_2)$ in Eq.(\ref{knu}) stands for the wave function of the 
final state (after photodetachment), while $\Psi_{{\rm H^{-}}}({\bf r}_1, {\bf r}_2)$ is the wave function of the ground $1^1S-$state of the hydrogen 
ion. In atomic units the last formula from Eq.(\ref{knu}) takes the form (compare with \cite{Bethe}, \cite{BS})
\begin{eqnarray}
 d\sigma_{\nu} = \alpha a^2_0 \frac{\mid {\bf p}_e \mid}{2 \pi \omega} \Bigl| \Bigl\{\int \int 
 \Psi^{*}_{fi}({\bf r}_1, {\bf r}_2) \Bigl[ {\bf e}_f \cdot \Bigl(\frac{\partial}{\partial {\bf r}_1} + \frac{\partial}{\partial 
 {\bf r}_2}\Bigr) \Bigr] \Psi_{{\rm H^{-}}}({\bf r}_1, {\bf r}_2) d^3{\bf r}_1 d^3{\bf r}_2 \Bigr\} \Bigr|^2 do \; \; \; \label{knu1} 
\end{eqnarray}
where $c$ is the speed of light in vacuum, $a_0$ is the Bohr radius and $v_a = \frac{e^2}{\hbar}$ is the atomic velocity. From here one finds that in atomic units
($a.u.$) the dimensionless ratio $\frac{c}{v_a}$ equals $\alpha^{-1} \approx 137$, where $\alpha$ is the fine structure constant. All other values in Eq.(\ref{knu1}) 
must be expressed in atomic units. Finally, one finds for the differential cross-section of the photodetachement
\begin{eqnarray}
 d\sigma_{\nu} = \frac{1}{\pi} \alpha a^2_0 \Bigl(\frac{p_e}{p^2_e + \gamma^2}\Bigr) \Bigl| \Bigl\{\int \int 
 \Psi^{*}_{fi}({\bf r}_1, {\bf r}_2) \Bigl[ {\bf e}_f \cdot \Bigl(\frac{\partial}{\partial {\bf r}_1} + \frac{\partial}{\partial 
 {\bf r}_2}\Bigr) \Bigr] \Psi_{{\rm H^{-}}}({\bf r}_1, {\bf r}_2) d^3{\bf r}_1 d^3{\bf r}_2 \Bigr\} \Bigr|^2 do \label{knu2}
\end{eqnarray}
where $p_e = \mid {\bf p}_e \mid$ and the factor $\frac12 \gamma^2$ equals to the ionization potential of the negatively charged hydrogen ion H$^{-}$. In these 
notations one finds $\hbar \omega = \frac12 p^2_e + \frac12 \gamma^2$, or $\omega = \frac12 p^2_e + \frac12 \gamma^2$ in atomic units (it was used in derivation of 
Eq.(\ref{knu2})). The formula, Eq.(\ref{knu}) - Eq.(\ref{knu2}), are used in numerical calculations of the differential and total cross-sections of photodetachment of 
the negatively charged H$^{-}$ ion. Note that the formula, Eq.(\ref{knu}), is written in the two forms: one-electron form and two-electron form. One-electron form is 
used in calculations, while two-electron form is needed to show the explicit invariance of the final expressions upon all possible electron-electron permutations. To 
explain all essential details of such calculations we need to derive the expressions which include the approximate wave functions of the bound $1^1S-$state in the 
H$^{-}$ ion(s) and wave functions of the final states, i.e. states which arise after photodetachment. Different forms of the final state wave functions explain 
different formulas which are used in the modern literature for these cross-sections. 

\section{Wave functions of the negatively charged hydrogen ions}

The wave functions of the ground $1^1S-$states in the hydrogen ions (${}^{\infty}$H$^{-}$, ${}^{3}$H$^{-}$ (or T$^{-}$), ${}^{2}$H$^{-}$ (or D$^{-}$) and 
${}^{1}$H$^{-}$) are approximated with the use of different variational expansions. One of the best such expansions is the exponential variational 
expansion in relative/perimetric coordinates coordinates. For the ground $1^1S-$state of the H$^{-}$ ion it takes the form \cite{Fro98}
\begin{eqnarray}
 \Psi_{{\rm H^{-}}}({\bf r}_1, {\bf r}_2) = \frac{1}{\sqrt{2}} \sum^{N}_{i=1} C_i \Bigl[ \exp(-\alpha_i r_{32} -\beta_i r_{31} - \gamma_i r_{21}) +  
  \exp(-\beta_i r_{32} -\alpha_i r_{31} - \gamma_i r_{21}) \Bigr] \label{exp}
\end{eqnarray}
where $\hat{P}_{12}$ is the permutation of the two identical particles 1 and 2 (electrons), the subscript 3 designates the hydrogen nucleus. In this 
notation $r_{21} = \mid {\bf r}_2 - {\bf r}_1 \mid$ is the electron-electron (or correlation) coordinate and $r_{3i} = \mid {\bf r}_3 - {\bf r}_i \mid$ are the 
corresponding electron-nuclear coordinates ($i = 1, 2$). Each of these three relative coordinates $r_{32}, r_{31}, r_{21}$ is a scalar which is rotationally and
translationally invariant. Also in this formula $C_i$ ($i$ = 1, $\ldots, N$) are the linear variational coefficients of the approximate (or trial) wave function. 
The $3 N-$parameters $\alpha_i, \beta_i, \gamma_i$ ($i$ = 1, $\ldots, N$) are the non-linear parameters (or varied parameters) of the trial wave function, 
Eq.(\ref{exp}). As follows from the results of our earlier studies (see, e.g., \cite{Fro2005}, \cite{Fro98}) the exponential variational expansion, Eq.(\ref{exp}), 
provides very high numerical accuracy for any bound state in arbitrary Coulomb three-body systems, i.e. for systems with arbitrary particle masses and electric 
charges. For the H$^{-}$ ions such an accuracy is much higher than accuracy which any other variational expansion may provide, if the same number of basis functions 
$N$ is used. Note also that from Eq.(\ref{exp}) one finds the following formula
\begin{eqnarray}
 {\bf e}_f \cdot \Bigl(\frac{\partial}{\partial {\bf r}_1} + \frac{\partial}{\partial {\bf r}_2}\Bigr) \Psi_{{\rm H^{-}}} =
 \Bigl({\bf e}_f \cdot \frac{{\bf r_{1}}}{r_{1}}\Bigr) \sum^{N}_{i=1} C_i \alpha_i f_i + \Bigl({\bf e}_f \cdot \frac{{\bf
 r_{2}}}{r_{2}}\Bigr) \sum^{N}_{i=1} C_i \beta_i f_i \\
  = \Bigl({\bf e}_f \cdot {\bf n}_{1}\Bigr) \Bigl[\frac{\partial}{\partial r_{1}} \Psi_{{\rm H^{-}}}\Bigr]
 + \Bigl({\bf e}_f \cdot {\bf n}_{2} \Bigr) \Bigl[\frac{\partial}{\partial r_{2}} \Psi_{{\rm H^{-}}}\Bigr] \nonumber
\end{eqnarray}
where ${\bf n}_i = \frac{{\bf r}_{3i}}{r_{3i}}$ is the unit vector of the $i$-th electron and $i$ = 1, 2. Note that in this expression all terms which contain 
the $\gamma_{i}$ parameters cancel each other, since we always have $\frac{\partial r_{21}}{\partial {\bf r}_{1}} = -\frac{\partial r_{21}}{\partial {\bf r}_{2}}$.

The presence of a large number of the varied non-linear parameters $\alpha_i, \beta_i, \gamma_i$ in Eq.(\ref{exp}) allows one to construct very compact, extremely 
flexible and highly accurate approximations to the actual wave functions of arbitrary three-body systems, including the negatively charged hydrogen ions. The 
results of our calculations obtained with the use of the exponential expansion Eq.(\ref{exp}), can be found in Table I. This Table contains only the total energies 
(in atomic units) and ionization potentials $\chi_1$ (in $eV$). In addition to the values shown in Table I we have determined a large number of expectation values 
which can be used to evaluate various bound state properties and probabilities of different processes/reactions in these ions. Some of these expectation values 
determined with our wave functions are shown in Tables II and III. The meaning of the notations used to designate some bound state properties is explained in
\cite{Fro98}. Table II illustrate actual convergence rates for different properties upon the total number of basis functions used. It allows us to predict the number 
of stable decimal digits in these and other similar bound state properties. Table III contains a large number of expectation values (only stable decimal digits are 
shown) determined in highly accurate computations. Note that the total non-relativistic energies and other expectation values obtained in our calculations are the most 
accurate values ever obtained for isotope substituted hydrogen ions: ${}^{3}$H$^{-}$ (or T$^{-}$), ${}^{2}$H$^{-}$ (or D$^{-}$) and ${}^{1}$H$^{-}$. Convergence of these
expectation values can be understood from Table IV. These expectation values indicate the overall quality of our wave functions used in calculations of the photodetachment 
cross-section of the H$^{-}$ ion(s). Another well known test for bound state wave functions of Coulomb few-body systems is numerical coincidence of the predicted and 
computed cusp values (see, e.g., \cite{Fro98}). Recently, we have found  that the quality of bound state wave functions can be tested in accurate evaluations of the 
lowest-order QED correction for the hydrogen ions (see Appendix I). Our values for the lowest-order QED corrections determined to high accuracy for the ${}^{\infty}$H$^{-}$, 
${}^{3}$H$^{-}$ (or T$^{-}$), ${}^{2}$H$^{-}$ (or D$^{-}$) and ${}^{1}$H$^{-}$ ions are presentd in Table V.   

It should be mentioned that in our calculations of the ${}^{3}$H$^{-}, {}^{2}$H$^{-}$ and ${}^{1}$H$^{-}$ hydrogen ions we have used the following updated 
values for the nuclear masses of the tritium, deuterium and protium (in $MeV/c^{2}$)
\begin{eqnarray}
 m_{e} = 0.510998910 \; \; \; , \; \; \; m_{p} = 938.272046 \\
 m_{d} = 1875.612859 \; \; \; , \; \; \; m_{t} = 2808.920906 \nonumber
\end{eqnarray}
In our calculations these masses are considered as exact. The masses of these three hydrogenic nuclei have recently been determined in high-energy experiments 
to better accuracy than they were known in the middle of 1990's. Usually, these masses are expressed in special high-energy mass units ($MeV/c^{2}$). Numerical 
value of the electron mass $m_e$ (or $m_e c^2$) can be used to re-calculate these masses to atomic units. Highly accurate computations of the ground states in the 
${}^{3}$H$^{-}$ (or T$^{-}$), ${}^{2}$H$^{-}$ (or D$^{-}$) and ${}^{1}$H$^{-}$ ions with these nuclear masses have never been performed.

\section{Final state wave functions}

After photodetachment of the H$^{-}$ ion, Eq.(\ref{e0}), in the final state we have the neutral hydrogen atom in one of its bound states and `free' electron which 
moves in the field of the hydrogen atom with the kinetic energy $E_e =  \frac{\hbar^2 k^2}{2 m_e}$, where $\hbar = \frac{h}{2 \pi}$ is the reduced Planck constant 
(or Dirac constant), $k$ is the wave number and $m_e$ is the electron mass. The corresponding wave function of the final state is represented as a product of the 
bound state wave function of the hydrogen atom and the wave function of the free electron. The wave function of the bound $(n \ell m)-$state of the hydrogen atom is 
written in the form (see, e.g., \cite{LLQ}) $\Phi_{n \ell m}(r, \Theta, \phi) = R_{n \ell}(Q, r) Y_{\ell m}(\Theta, \phi)$, where $Y_{\ell m}(\Theta, \phi) = 
Y_{\ell m}({\bf n})$ is a spherical harmonic and $R_{n \ell}(Q, r)$ is the one-electron radial function. The radial function $R_{n \ell}(Q, r)$ takes the form (see, 
e.g., \cite{LLQ})
\begin{eqnarray}
  R_{n \ell}(Q,r) = \frac{1}{r n} \sqrt{\frac{Q (n - \ell - 1)!}{(n +
  \ell)!}} \Bigl[ \frac{2 Q r}{n} \Bigr]^{\ell + 1} \Bigl\{ 
  \sum^{n-\ell-1}_{k=0}
  \frac{(-1)^k}{k!}
  \left(
  \begin{array}{c}
   n + \ell \\
   2 \ell + k + 1
  \end{array}
  \right)
  \Bigl[ \frac{2 Q r}{n} \Bigr]^{k} \Bigr\} \times \nonumber \\
 \exp\Bigl(-\frac{Q r}{n}\Bigr) \label{hydrogen}
\end{eqnarray}
where $Q$ (= 1) is the nuclear charge, while $n$ and $\ell$ are the quantum numbers of this bound state. Note that all radial functions defined by Eq.(\ref{hydrogen}) 
have unit norms. The six following functions of the hydrogen atom are very important for calculations of the photodetachment cross-sections:
\begin{eqnarray}
  R_{10}(r) = 2 \exp(- r) \; \; \; , \; \; \; R_{20}(r) = \frac{1}{\sqrt{2}} \exp(- \frac{r}{2}) \Bigl( 1 - \frac{r}{2} \Bigr) \; \; \; , \; \; \; 
  R_{21}(r) = \frac{1}{2 \sqrt{6}} r \exp(- \frac{r}{2}) \label{hydr} \\
  R_{30}(r) = \frac{2}{3 \sqrt{3}}  \Bigl( 1 - \frac{2 r}{3} + \frac{2 r^2}{27} \Bigr) \exp(- \frac{r}{3}) \; \; \; , \; \; \;  
  R_{31}(r) = \frac{8}{27 \sqrt{6}}  \Bigl( 1 - \frac{r}{6} \Bigr) \exp(- \frac{r}{3}) \nonumber
\end{eqnarray}
The final states with these hydrogen wave functions provide largest numerical contributions in the photodetachment cross-section. Note also that in this study all angular 
parts of these hydrogenic functions (spherical harmonics) are included in the formulas for matrix elements (see, e.g., Eq.(\ref{int0}) below). 

The wave function of the free electron is represented as the sum of the plane wave $\phi({\bf r}) = \exp(\imath {\bf k} \cdot {\bf r})$ and converging spherical waves 
which are usually designated as $\phi^{(-)}_{{\bf p}}({\bf r})$ \cite{LLQ}. In the non-relativistic dipole approximation used here the following selection rule is 
applied: transitions from the incident $S(L = 0)-$state of the H$^{-}$ ion can only be into the final $P(L = 1)-$state of the two-body (final) system: the H-atom plus 
free electron. It follows from here that in the final states we can detect the H-atom in one of its $s(\ell = 0)-$states and free electron which moves away in the 
$p$-wave. This is the $sp-$channel of the H$^{-}$ ion photodetachment which produces the largest contribution into photodetachment cross-section of the H$^{-}$ ion. 
Another possibility is the formation of the final H-atom in one of its three $p(\ell = 1)-$states and free electron which moves away in the $s-$wave. This represents 
the $ps-$channel in the photodetachment of the H$^{-}$ ion. 

The wave function of the final state for the $sp-$channel takes the form (see, e.g., \cite{AB})
\begin{eqnarray}
 \psi_f(1,2) = R_{n 0}(Q = 1, r_{31}) ({\bf n}_e \cdot {\bf n}_{2}) \Phi_{p 1}(r_{32}) = ({\bf n}_e \cdot {\bf n}_2) R_{n 0}(r_{31}) j_1(p_e r_{32}) \label{hfinal}
\end{eqnarray}
where ${\bf n}_2 = \frac{{\bf r}_{32}}{r_{32}}$ is the unit vector of the second (unbound) electron and ${\bf n}_e = \frac{{\bf p}_e}{p_e}$ is the direction of the 
momentum of the outgoing photo-electron ${\bf p}_e$, i.e. the second electron which becomes unbound after the photodetachment of the hydrogen negatively charged ion. In 
Section III this vector has been designated by ${\bf p}$, i.e. ${\bf p}_e = {\bf p}$. The notation $R_{n \ell=0}(Q = 1, r_{31}) = R_{n 0}(r_{31})$ stands for the radial 
function of the bound state of the hydrogen atom with quantum numbers $n \ge 1$ and $\ell = 0$. In this equation the notation $\Phi_{p_e 1}(r_{32}) = j_1(p_e r_{32})$ 
stands for the wave function of the electron which becomes `free' after photodetachment. Here and below $j_1(x)$ and $j_0(x)$ are the spherical Bessel functions of the 
first and zero-orders, respectively. Such a wave function represents an electron which moves in the field of the neutral hydrogen atom. This electron has an angular 
momentum $\ell = 1$ and absolute momentum $p_e = \mid {\bf p}_e \mid$.

The wave function of the final state for the $ps-$channel can be written in the following form
\begin{eqnarray}
 \psi_f(1,2) = R_{n \ell=1}(Q = 1, r_{31}) Y_{1m}({\bf n}_1) \frac{1}{2 p_e} \Phi_{p \ell=0}(r_{32}) = \frac{1}{2 p_e} {\bf n}_{1m} \cdot R_{n 1}(r_{31}) j_0(p_e r_{32}) 
 \label{arl}
\end{eqnarray}
where $R_{n 1}(Q = 1, r_{31}) = R_{n 1}(r_{31})$ is the radial function of the bound $\mid n 1 \rangle$-state of the hydrogen atom with the principal quantum number$n$ ($n 
\ge 2$) and angular quantum number $\ell = 1$. Also, in Eq.(\ref{arl}) the notation $Y_{1m}({\bf x})$ stands for the corresponding spherical harmonics \cite{LLQ}, while 
the vector ${\bf n}_{1} = \frac{{\bf r}_{31}}{r_{31}}$ has the following components: $({\bf n}_{1})_{+1} = - \frac{1}{\sqrt{2}} (n_x + \imath n_y), ({\bf n}_{1})_{0} = n_z, 
({\bf n}_{1})_{+1} = - \frac{1}{\sqrt{2}} (n_x - \imath n_y)$, where $n_x, n_y, n_z$ are three Cartesian components of the unit vector ${\bf n}_1$. Note that all 
calculations for the $ps-$channel in the photodetachment of the H$^{-}$ ion are significantly more complicated than for the $sp-$channel. One obvious complication can 
directly be seen from Eq.(\ref{arl}), where the final wave function is a vector. In this study we restrict ourselves to the consideraion of the `traditional' $sp-$channel 
only. It is clear $a$ $priori$ that the overall contributions of all $ps-$channels into photodetachment of the H$^{-}$ ions in actual stars is very small. Excitation 
energies which are needed to form the final hydrogen atoms in the $np$-states are extremely high ($\Delta E \ge$ 10.2043 $eV$). The corresponding `equilibrium' temperatures 
at the stellar surface are above $10^5$ $K$ (they correspond to very hot O-stars also known as the Wolf-Rayet stars). However, as mentioned in the Introduction for such stars 
photodetachment of the H$^{-}$ ions dos not play any role.  

\section{Formulas for the matrix elements}

In this Section we derive analytical expressions for the photodetachment cross-section(s). First, let us discuss the angular integration in the expression for $M_{i \rightarrow f}$ 
included into Eq.(\ref{knu1}). Below we restrict ourselves to the case of the $sp-$channel only. In this case the final state wave function is represented in the form 
of Eq.(\ref{hfinal}) and the formula for the $M_{i \rightarrow f}$ amplitude is written in the following form
\begin{eqnarray}
 & & M_{i \rightarrow f} = \int \int \Bigl[ A ({\bf p}_e \cdot {\bf n}_{32}) ({\bf e}_f \cdot {\bf n}_{31}) B_1 + A ({\bf p}_e \cdot {\bf n}_{32}) ({\bf e}_f \cdot {\bf n}_{32}) 
 B_2 \Bigr] d{\bf r}_{31} d{\bf r}_{32} \label{ee1} \\ 
 &=& \int \int \oint \oint \Bigl[ A ({\bf p}_e \cdot {\bf n}_{32}) ({\bf e}_f \cdot {\bf n}_{31}) B_1 + A ({\bf p}_e \cdot {\bf n}_{32}) ({\bf e}_f \cdot {\bf n}_{32}) 
 B_2 \Bigr] r^{2}_{31} dr_{31} r^{2}_{32} dr_{32} d{\bf n}_{31} d{\bf n}_{32} \nonumber
\end{eqnarray}
where $A, B_{1}$ and $B_{2}$ are some scalar expressions (see below). Result of angular integration of the first term in the last equation equals zero identically, since its integrand 
is a scalar upon ${\bf n}_{31}$. The integration of the second term produces the factor $\frac{4 \pi}{3} ({\bf e}_f \cdot {\bf n}_{e})$. By introducing the explicit expressions for 
the $A, B_{1}$ and $B_{2}$ scalar values we obtain the following formula for the transition amplitude $M_{i \rightarrow f}$:
\begin{eqnarray}
 & & M_{i \rightarrow f} = 4 \pi \imath ({\bf n}_e \cdot {\bf e}_f) \sum^{N}_{j=1} C_j \Bigl[\alpha_j \int_{0}^{+\infty} 
 \int_{0}^{+\infty} \int^{r_{31} + r_{32}}_{\mid r_{31} - r_{32} \mid} R_{n0}(r_{31}) j_{1}(p_e r_{32}) \times \nonumber \\
 & & \exp(-\alpha_j r_{32} -\beta_j r_{31} - \gamma_j r_{21}) r_{32} r_{31} r_{21} dr_{32} dr_{31} dr_{21} + \beta_j \times \label{int0} \\
  & & \int_{0}^{+\infty} \int_{0}^{+\infty} \int^{r_{31} + r_{32}}_{\mid r_{31} - r_{32} \mid} R_{n0}(r_{31}) j_{1}(p_e r_{32})
 \exp(-\beta_j r_{32} -\alpha_j r_{31} - \gamma_j r_{21}) r_{32} r_{31} r_{21} dr_{32} dr_{31} dr_{21} \Bigr] \nonumber
\end{eqnarray}
where the notation $\Phi_{\ell = 1}(p_e; r_2)$ designates the wave function of the `free' electron which moves in the field of the neutral hydrogen atom. As one can see from 
Eq.(\ref{int0}) numerical calculations of the transition amplitude $M_{i \rightarrow f}$ are reduced to the calculation of the three-body integrals in relative coordinates $r_{32}, 
r_{31}, r_{21}$. Calculations of such integrals are discussed in detail in the next Section. Here we want to note that the photodetachment cross-section for the $sp-$channel always 
contains the factor $\mid M_{i \rightarrow f} \mid^2$ which is proportional to the factor $({\bf p}_e \cdot {\bf e})^2 = p^2_e ({\bf n}_e \cdot {\bf e})^2$ (here and below ${\bf e} 
= {\bf e}_f$).

To determine the factor $({\bf p}_e \cdot {\bf e})^2 = p^2_e({\bf n}_e \cdot {\bf e})^2$ we note that in 99.99 \% of all stellar photospheres the hydrogen ions are free-oriented. This 
means that the expression from Eq.(\ref{ee1}) must be averaged over possible directions of the unit vectors ${\bf e}_1$ and ${\bf e}_2$ which describe the polarization of the incident 
photon. In the incident light wave we always have ${\bf e}_{1} \perp {\bf e}_{2} \perp {\bf k}_f$, where the unit vector ${\bf k}_f$ corresponds to the direction of propagation of this 
light wave, or the corresponding light quantum (photon). An arbitrary scalar expression can be averaged over all possible directions of the unit vectors ${\bf e}_1$ and ${\bf e}_2$ 
with the use of the following formulas
\begin{equation}
 \overline{e_i e_k} = \frac12 (\delta_{ik} - (k_f)_{i} (k_f)_{k})
\end{equation}
and
\begin{equation}
 \overline{({\bf a} \cdot {\bf e}) ({\bf b} \cdot {\bf e})} = \frac12 [({\bf a} \cdot {\bf b}) - ({\bf a} \cdot {\bf k}_f) ({\bf b} \cdot
 {\bf k}_f)] = \frac12 ({\bf a} \times {\bf k}_f) ({\bf b} \times {\bf k}_f) \label{aver}
\end{equation}
where ${\bf a}$ and ${\bf b}$ are the two arbitrary vectors and the notation ${\bf x} \times {\bf y}$ stands for the vector product of the two vectors ${\bf x}$ and 
${\bf y}$. From Eq.(\ref{aver}) one finds that 
\begin{equation}
 \overline{({\bf n}_e \cdot {\bf e})^2} = \frac12 ({\bf n}_e \times {\bf k}_f)^2
\end{equation}
Now, the explicit formula for the differential cross-section, Eq.(\ref{knu2}), takes the form 
\begin{eqnarray}
 d\sigma_{\nu} = 8 \pi \alpha a^2_0 {\cal K} \Bigl(\frac{p_e}{p^2_e + \gamma^2}\Bigr) ({\bf n}_e \times {\bf k}_f)^2 \Bigl| {\cal F}_{i \rightarrow f} \Bigr|^2 do
  \label{sigmtot} 
\end{eqnarray}
where ${\cal K}$ is some additonal (numerical) factor and the matrix element ${\cal F}_{i \rightarrow f}$ (photodetachment amplitude) is written in the form 
\begin{eqnarray}
 & & {\cal F}_{i \rightarrow f} = \sum^{N}_{j=1} C_j \Bigl[\alpha_j \int_{0}^{+\infty} \int_{0}^{+\infty} \int^{r_{31} + r_{32}}_{\mid r_{31} - r_{32} \mid} R_{n0}(r_{31}) 
 j_{1}(p_e r_{32}) \times \nonumber \\
 & & \exp(-\alpha_j r_{32} -\beta_j r_{31} - \gamma_j r_{21}) r_{32} r_{31} r_{21} dr_{32} dr_{31} dr_{21} + \beta_j \times \label{finalamp} \\
  & & \int_{0}^{+\infty} \int_{0}^{+\infty} \int^{r_{31} + r_{32}}_{\mid r_{31} - r_{32} \mid} R_{n0}(r_{31}) j_{1}(p_e r_{32})
 \exp(-\beta_j r_{32} -\alpha_j r_{31} - \gamma_j r_{21}) r_{32} r_{31} r_{21} dr_{32} dr_{31} dr_{21} \Bigr] \nonumber
\end{eqnarray}
The factor ${\cal K}$ is responsible for missing normalization constants in the incident and final wave functions. The same factor includes one part of the finite mass corrections for the 
protium, deuterium and tritium (hydrogen isotopes). Another part of such a correction comes from the photodetachment amplitude ${\cal F}_{i \rightarrow f}$. 

As follows from Eq.(\ref{sigmtot}) the photodetachment cross-section is always proportional to the factor $({\bf n}_e \times {\bf k}_f)^2 = \sin^2 \Theta$, where $\Theta$ is the angle 
between the directions of propagation of the incident photon ${\bf k}_f$ and final electron ${\bf n}_e$ (or photo-electron). This results corresponds to the well known dipole approximation.
As follows from Eq.(\ref{finalamp}) the matrix element ${\cal F}_{i \rightarrow f}$ does not depend upon any of the angular variables. Therefore, the formula for the total cross-section of
the photodetachment takes the form
\begin{eqnarray}
 \sigma_{\nu} = \frac{64 \pi^2}{3} \alpha a^2_0 {\cal K} \Bigl(\frac{p_e}{p^2_e + \gamma^2}\Bigr) \Bigl| {\cal F}_{i \rightarrow f} \Bigr|^2 \label{sigmtot1} 
\end{eqnarray} 
Now, we need to perform the last step of our procedure and calculate three-electron integrals in the relative coordinates. Note that each of these integrals contains one spherical Bessel 
function $j_{q}(p r_{32})$, i.e. $j_{1}(p r_{32})$ for the $sp-$channel and $j_{0}(p r_{32})$ for the $ps-$channel of the H$^{-}$ photodetachment. 
   
\section{Three-body integrals in relative coordinates}

As we have shown above the problem of analytical and numerical computations of matrix elements arising in Eq.(\ref{knu}) and Eq.(\ref{int0}) with the trial wave function defined in Eq.(\ref{exp}) 
is reduced to the computation of the following three-body integrals with the Bessel functions $j_{0}(k r_2) = j_{0}(p r_2)$ for the $ps-$channel and $j_{1}(k r_2) = j_{1}(p r_2)$ for the 
$sp-$channel \cite{FrWa}
\begin{eqnarray}
 \int_{0}^{+\infty} \int_{0}^{+\infty} \int^{r_{31} + r_{32}}_{\mid r_{31} - r_{32} \mid} \phi_{n\ell}(r_{31}) j_{q}(p r_{32}) \exp(-\alpha r_{32} -\beta r_{31} - 
 \gamma r_{21}) r_{32} r_{31} r_{21} dr_{32} dr_{31} dr_{21} \label{int1}
\end{eqnarray}
where $q$ = 0, 1 and $r_{21} = \mid {\bf r}_2 - {\bf r}_1 \mid$ is the electron-electron coordinate which varies between the following limits $\mid r_{32} - r_{31} \mid 
\le r_{21} \le r_{32} + r_{31}$. In Eq.(\ref{int1}) the notation $\phi_{n\ell}(r_{31})$ stands for the unit norm wave functions $R_{n0}(r_{31})$ and $R_{n1}(r_{31})$ of 
the bound states of the hydrogen atom, Eqs.(\ref{hydrogen}) and (\ref{hydr}). All other notations in Eq.(\ref{int1}) are defined above in Eq.(\ref{exp}). The integrals 
defined by Eq.(\ref{int1}) are computed analytically in perimetric coordinates $u_1, u_2, u_3$, where $u_1 = \frac12 (r_{31} + r_{21} - r_{32}), u_2 = \frac12 (r_{32} + r_{21} - 
r_{31})$ and $u_3 = \frac12 (r_{31} + r_{32} - r_{21})$. The three perimetric coordinates $u_1, u_2, u_3$ are independent of each other and each of them varies between 
0 and +$\infty$.

Let us derive the closed analytical expression for the integrals, Eq.(\ref{int1}) with $q = 0$ and $q = 1$. First, we need to obtain the general formula for the closely 
related auxiliary three-body integral $\Gamma_{k,l,n}(\alpha,\beta,\gamma)$ which is written in the form
\begin{eqnarray}
 \Gamma_{k,l,n}(\alpha,\beta,\gamma) =
 \int_{0}^{+\infty} \int_{0}^{+\infty} \int^{r_{31} + r_{32}}_{\mid r_{31} - r_{32} \mid} r^{k}_{32} r^{l}_{31} r^{n}_{21} \exp(-\alpha r_{32} -\beta r_{31} - \gamma r_{21}) 
  dr_{32} dr_{31} dr_{21} \label{int2}
\end{eqnarray}
where $k \ge 0, l \ge 0, n \ge 0$ and $\alpha + \beta > 0, \alpha + \gamma > 0$ and $\beta + \gamma > 0$ (see below). Analytical computation of this integral has extensively been 
explained in a number of erlier studies. Correspondingly, below we restrict ourselves only to a few following remarks. In perimetric coordinates the integral, Eq.(\ref{int2}), 
takes the form
\begin{eqnarray}
 \Gamma_{k,l,n}(\alpha,\beta,\gamma) = 2 \int_0^{\infty} \int_0^{\infty} \int_0^{\infty} exp\Bigl[-(\alpha + \beta) u_3 -(\alpha + \gamma) u_2
 -(\beta + \gamma) u_1\Bigr] (u_2 + u_3)^k \nonumber \\
(u_1 + u_3)^l (u_1 + u_2)^n du_1 du_2 du_3 \label{int3}
\end{eqnarray}
where we took into account the fact that the Jacobian of transformation from the relative $(r_{32}, r_{31}. r_{21})$ to perimetric coordinates $(u_1, u_2, u_3)$ equals 2. The integration over 
three independent perimetric coordinates $u_i$ ($0 \le u_i < \infty$) in Eq.(\ref{int3}) is simple and explicit formula for the $\Gamma_{n,k,l}(\alpha,\beta,\gamma)$ integral is reduced to the 
form
\begin{eqnarray}
 & &\Gamma_{k;l;n}(\alpha, \beta, \gamma) = 2 \sum^{k}_{k_1=0} \sum^{l}_{l_1=0} 
 \sum^{n}_{n_1=0} C_{k}^{k_1} C_{l}^{l_1} C_{n}^{n_1} 
 \frac{(l-l_1+k_1)!}{(\alpha + \beta)^{l-l_1+k_1+1}}
 \frac{(k-k_1+n_1)!}{(\alpha + \gamma)^{k-k_1+n_1+1}}
 \frac{(n-n_1+l_1)!}{(\beta + \gamma)^{n-n_1+l_1+1}} \nonumber \\
 &=& 2 \cdot k! \cdot l! \cdot n! \sum^{k}_{k_1=0} \sum^{l}_{l_1=0} \sum^{n}_{n_1=0} 
 \frac{C^{k_1}_{n-n_1+k_1} C^{l_1}_{k-k_1+l_1} C^{n_1}_{l-l_1+n_1}}{(\alpha + 
 \beta)^{l-l_1+k_1+1} (\alpha + \gamma)^{k-k_1+n_1+1} (\beta + \gamma)^{n-n_1+l_1+1}} 
 \label{int4}
\end{eqnarray}
where $C^{m}_{k}$ are the binomial coefficients (= number of combinations from $k$ by $m$) (see, e.g., \cite{GR}). In many cases all three integer numbers $k, l, m$ are relatively small ($\le 3$) 
and this simplifies numerical applications of the formula, Eq.(\ref{int4}).

The parameters $\alpha, \beta$ and $\gamma$ in Eqs.(\ref{int1}) - (\ref{int4}) can be arbitrary real and/or complex numbers. Furthermore, one of these three parameters can be equal zero, but the 
three principal conditions for these parameters $\alpha + \beta > 0, \alpha + \gamma > 0$ and $\beta + \gamma > 0$ must always be obeyed. If these three parameters are complex, then these three 
conditions are written in the form $Re(\alpha + \beta) > 0, Re(\alpha + \gamma) > 0$ and $Re(\beta + \gamma) > 0$. These conditions are needed to guarantee convergence of the auxiliary three-body 
integrals $\Gamma_{k;l;n}(\alpha, \beta, \gamma)$, Eq.(\ref{int2}).

Now, we can derive the explicit formulas for the analogous three-body integrals which contain the Bessel functions $j_{0}(p r_2)$ and/or $j_{1}(p r_2)$, where $p = p_e$. First, let us obtain the 
computational formula for the integral which contains the $j_0(p r_{32})$ function
\begin{eqnarray}
 B^{(0)}_{k;l;n}(\alpha, \beta, \gamma; p) &=& \int_{0}^{+\infty} \int_{0}^{+\infty} \int^{r_{32} + r_{31}}_{\mid r_{32} - r_{31} \mid} r^{k}_{32} r^{l}_{31} r^{n}_{21} j_0(p r_{32}) \times 
\nonumber \\ 
 & & \exp(-\alpha r_{32} - \beta r_{31} - \gamma r_{21}) dr_{32} dr_{31} dr_{21} \label{e33}
\end{eqnarray}
By using the formula $j_0(x) = \frac{\sin x}{x}$ one finds
\begin{eqnarray}
 B^{(0)}_{k;l;n}(\alpha, \beta, \gamma; p) = \sum^{\infty}_{q=0} \frac{(-1)^q p^{2 q}}{(2 q + 1)!}  \Gamma_{k + 2 q;l;n}(\alpha, \beta, \gamma) \approx \sum^{q_{max}}_{q=0} \frac{(-1)^q 
 p^{2 q}}{(2 q + 1)!} \Gamma_{k + 2 q;l;n}(\alpha, \beta, \gamma) \label{e34}
\end{eqnarray}
where $\Gamma_{k;l;n}(\alpha, \beta, \gamma)$ is the integral defined in Eq.(\ref{int3}).

The integral $B^{(0)}_{k;l;n}(\alpha, \beta, \gamma; p)$ in the last equation converges for relatively small $p$, but for $p \le 1$ it converges very rapidly. In reality, the maximal values of the 
index $q$ (i.e. $q_{max}$) in Eq.(\ref{e34}) are always finite. Numerical investigations indicate that to stabilize 15 decimal digits for $k \le 1$ one needs to use in Eq.(\ref{e34}) $q_{max}$ = 5 
and $q_{max} = 10$. For $p \ge 2$ the value of $q_{max}$ rapidly increases up to 40 - 50. The same conclusion is true about the convergence of the three-body integrals with the Bessel function 
$j_1(x) = \frac{\sin x}{x^2} - \frac{\cos x}{x}$. This integral takes the form
\begin{eqnarray}
 B^{(1)}_{k;l;n}(\alpha, \beta, \gamma; p) &=& \int_{0}^{+\infty} \int_{0}^{+\infty} 
 \int^{r_{32} + r_{31}}_{\mid r_{32} - r_{31} \mid} r^{k}_{32} r^{l}_{31} r^{n}_{21} j_{1}(p r_{32}) 
 \exp(-\alpha r_{32} - \beta r_{31} - \gamma r_{21}) dr_{32} dr_{31} dr_{21} \nonumber \\
 &=& \sum^{\infty}_{q=0} \frac{(-1)^q (2 q + 2) 
 p^{2 q + 1}}{(2 q + 3)!} \Gamma_{k + 2 q + 1;l;n}(\alpha, \beta, \gamma) \label{e35}
\end{eqnarray}
The knowledge of the three-body integrals with the Bessel functions $j_0(x)$ and $j_1(x)$ allow one to determine all integrals, Eq.(\ref{int1}), needed in calculations of the photodetachment 
cross-section of the negatively charged hydrogen ion. The integrals $B^{(1)}_{k;l;n}(\alpha, \beta, \gamma; p)$ are needed to calculate the photodetachment cross-sections in the $sp-$channel, 
while the integrals $B^{(0)}_{k;l;n}(\alpha, \beta, \gamma; p)$ are needed to evaluate analogous cross-sections in the $ps-$channel. 

Finally, let us present the formulas for the three-body integrals with the spherical Bessel functions $j_0(p_e r_{31})$ and $j_1(p_e r_{32})$ and a few radial hydrogenic functions $R_{10}(r), 
R_{20}(r), R_{21}(r), R_{31}(r)$, where 
\begin{eqnarray}
  R_{10}(r) &=& 2 Q \sqrt{Q} exp(-Q r) \; \; \; , \; \; \; R_{20}(r) = \frac{Q \sqrt{Q}}{\sqrt{2}} \Bigl( 1 - \frac12 Q r \Bigr) exp(-\frac{Q r}{2}) \nonumber \\
 R_{21}(r) &=& \frac{Q \sqrt{Q}}{\sqrt{24}} Q r exp(-\frac{Q r}{2}) \; \; \; , \; \; \; 
 R_{31}(r) = \frac{8 Q \sqrt{Q}}{27 \sqrt{6}} \Bigl( 1 - \frac16 Q r \Bigr) exp(-\frac{Q r}{3}) \nonumber
\end{eqnarray}
where $Q = 1$ for the hyrogen atom. For the photodetachment of the hydrogen negative ion only the integrals which contain the following products $R_{10}(r_{31}) j_{1}(p r_{32}), R_{20}(r_{31}) 
j_{1}(p r_{32})$ and $R_{21}(r_{31}) j_{0}(p r_{32}), R_{31}(r_{31}) j_{0}(p r_{32})$ are important. As follows from Eqs.(\ref{e34}) and (\ref{e35}) these integrals are
\begin{eqnarray}
  & & 2 Q \sqrt{Q} B^{(1)}_{1;1;1}(\alpha, \beta + Q, \gamma; p) \; \; \; , \\  
  & & \frac{Q \sqrt{Q}}{2} B^{(1)}_{1;1;1}(\alpha, \beta + \frac{Q}{2}, \gamma; p) - \frac{Q \sqrt{Q}}{4} B^{(1)}_{1;2;1}(\alpha, \beta + \frac{Q}{2}, \gamma; p)
\end{eqnarray}
and 
\begin{eqnarray}
 & & \frac{Q^2 \sqrt{Q}}{24}  B^{(0)}_{1;2;1}(\alpha, \beta + \frac{Q}{2}, \gamma; p) \; \; \; , \\
 & & \frac{8 Q \sqrt{Q}}{27 \sqrt{6}} B^{(0)}_{1;1;1}(\alpha, \beta + \frac{Q}{3}, \gamma; p) - \frac{4 Q^2 \sqrt{Q}}{81 \sqrt{6}} B^{(0)}_{1;2;1}(\alpha, \beta + \frac{Q}{3}, \gamma; p) \; \; \; ,
\end{eqnarray}
respectively. Analogous formulas can be written for the integrals which include other radial hydrogenic functions, e.g., $R_{n0}(r)$ and $R_{m1}(r)$, where $n \ge 3$ and $m \ge 4$. However, the 
overall contribution of such integrals to the photodetachment cross-section of the H$^{-}$ ion ($Q = 1$ in all formulas above) is very small and they can be neglected in the lowest order 
approximation.

To test our formulas we perform calculations of the photodetachment cross-sections of the ${}^{\infty}$H$^{-}$, ${}^{3}$H$^{-}$ (or T$^{-}$), ${}^{2}$H$^{-}$ (or D$^{-}$) and ${}^{1}$H$^{-}$ ions. 
These cross-sectons are determined to high accuracy and for a large number of photo-electron momenta/energies (see Tables VI - VIII). It is shown in \cite{Fro2014} that the additional factor ${\cal 
K}$ in Eqs.(\ref{sigmtot}) and (\ref{sigmtot1}) equals $\frac{1}{8 \pi^2}$ which is the normalization constant of the angular part of the three-particle (or two-electron) wave functions of the H$^{-}$
ions. This leads to the following formula for the total cross-section of photodetachment
\begin{eqnarray}
 \sigma_{\nu} = \frac{8}{3} \alpha a^2_0 \Bigl(\frac{p_e}{p^2_e + \gamma^2}\Bigr) \Bigl| {\cal F}_{i \rightarrow f} \Bigr|^2 \label{sigmtt} 
\end{eqnarray} 
for the ${}^{\infty}$H$^{-}$ ion and 
\begin{eqnarray}
 \sigma_{\nu} \approx \frac{8}{3} \alpha a^2_0 \Bigl(1 + \frac{1}{M+1}\Bigr)^{2} \Bigl[\frac{p_e}{p^2_e (1 + \frac{1}{M+1}) + \gamma^2}\Bigr] \cdot \Bigl| {\cal F}_{i \rightarrow f} \Bigr|^2 \label{sigmttf} 
\end{eqnarray}
for the hydrogen isotopes with finite nuclear masses. In Eq.(\ref{sigmttf}) the notation $M$ stands for the mass of the nucleus of the corresponding hydrogen isotope (in atomic units). Results of 
our computations of the photodetachment cross-sections of the hydrogen negatively charged ions can be found in Tables VI, VII and VIII. In these calculations we have used our variational wave functions 
with $N$ = 300 and $N$ = 350 for the ${}^{\infty}$H$^{-}$ ion and $N$ = 350 for other hydrogen isotopes ($N$ is the total number of basis functions used). The corresponding total energies for each trial 
wave function (in atomic units) are given in Table VI. For the ${}^{\infty}$H$^{-}$ ion our photodetachment cross-section is comparable with the analugous cross-section from \cite{PRA1994} (only our 
cross-section is more accurate). Comparison with other papers where the photodetachment cross-section of the negatively charged hydrogen ${}^{\infty}$H$^{-}$ ion has been determined in the velocity 
representation (see references in \cite{PRA2014}) indicate clearly that our photodetachment cross-section (for this ion) are in good agreement with the cross-sections known from earlier works. Note again 
that the ${}^{\infty}$H$^{-}$ ion is a model atomic (or two-electron) system with the infinite nuclear mass. To the best of our knowledge the photodetachment cross-sections of the hydrogen ions with finite 
nuclear masses have never been determined in earlier works. In reality photodetachment of the negatively charged hydrogen ions with the finite nuclear masses requires an additional consideration. In our 
current approach we just introduced the corresponding mass-dependent corrections in the normalization factors of the hydrogenic wave functions (for the final hydrogen atom). All numerical calculations have
been performed by using the formulas Eqs.(\ref{sigmtt}) - (\ref{sigmttf}). A large number of additional results for the photodetachment cross-sections of the negatively charged hydrogen ions will be 
presented in our next paper \cite{Fro2014}, which inlcudes the photodetachment cross-sections in those cases when the final hydrogen atom is formed in the excited $s-$states. 
    
\section{A simple method to evaluate cross-sectioni}

The photodetachment cross-section of the negatively charged H$^{-}$ ion in the $sp-$channel can approximately be evaluated with the use of the following asymptotic method. Let us designate the 
total number of bound electrons in atomic system by $N_e$, while the notation $Q$ stands below for the nuclear charge $Q$. As is well known (see, e.g., \cite{BhaDra} - \cite{FroSm03} and 
references therein) the long range, single-electron asymptotic of the wave function of $N_e-$electron atomic system with the nuclear charge $Q$ is written in the form
\begin{equation}
 \mid \Psi(r) \mid = C r^{\frac{Z}{t} - 1} \exp(-t r) \label{eq32}
\end{equation}
where $C$ is some numerical constant, $Z = Q - N_e + 1$, while $t = \sqrt{2 I_1}$ and $I_1 = \chi_1$ is the first ionization potential which corresponds to the dissociation process H$^{-}$ = 
H + $e^{-}$. For the H$^{-}$ ion one finds in Eq.(\ref{eq32}): $Q = 1, N_e = 2$, and therefore, $Z = 0$ and the long range asymptotic of the wave function, Eq.(\ref{eq32}), is represented in 
the Yukawa-type form $\mid \Psi(r) \mid = \frac{C}{r} \exp(-t r)$, where $C$ is a constant which must provide the best correspondence with the highly accurate wave function of the H$^{-}$ ion 
at large $r$. The highly accurate wave function of the H$^{-}$ ion is assumed to be known. The explicit formula which is used to determine the constant $C$ is $C = \mid \Psi_{{\rm H}^{-}}(r) 
\mid r \exp(t r)$, where $\Psi_{{\rm H}^{-}}(r) = \Psi_{{\rm H}^{-}}(r_{32} = r, r_{31} = 0, r_{21} = r)$. As follows from the results of actual computations the factor $C$ in Eq.(\ref{eq32}) 
does not change for relatively large interval of variations of $r$, e.g., for $r$ bounded between 7.5 and 35 atomic units. The determined numerical value of this constant $C$ for the negatively 
charged hydrogen ions varies between $\approx 0.3562404$ (${}^{\infty}$H$^{-}$) and $\approx 0.3565232$ (${}^{1}$H$^{-}$). These values are larger than the corresponding values obtained in 
\cite{FroSm03} by a factor $2 \sqrt{2}$. It is important to note that this simple asymptotic method is essentially a single-electron approach. Therefore, only three Cartesian coordinates ${\bf 
r}$ and one radial variable $r$ are used below. These coordinates correspond to the electron which was originally bound to the neutral hydrogen atom. At the final stage of the process this 
electron becomes free.  

The photodetachment cross-section $\sigma(H^-)$ of the H$^-$ ion is written in the form which is similar to Eq.(\ref{knu2}) (see, e.g., \cite{John})   
\begin{eqnarray}
 d\sigma({\rm H}^-) = \alpha a^{2}_{0} \frac{p_e}{2 \pi \omega} \mid {\bf e}_f \cdot \int \psi^{*}_f \Bigl(\frac{\partial}{\partial {\bf r}} \psi_{i}\Bigr) d^3{\bf r} \mid^2 do
 = \alpha a^{2}_{0} \frac{p_e}{2 \pi \omega} \mid {\bf e}_f \cdot \int \psi_i \Bigl(\frac{\partial}{\partial {\bf r}} \psi_{f}\Bigr) d^3{\bf r} \mid^2 do \; \; \; \label{cr-sec}
\end{eqnarray}
The normalized wave function of the incident state is $C r^{\frac{Z}{t} - 1} \exp(-t r) = \frac{C}{r} \exp(-t r)$, while the wave function of the final state is $\exp( \imath {\bf p}_e \cdot {\bf 
r})$. By substituting these exressions into Eq.(\ref{cr-sec}) one finds
\begin{eqnarray}
 d\sigma({\rm H}^-) &=& \alpha a^{2}_{0} \Bigl( \frac{16 \pi^2 p^{3}_e}{2 \pi \omega} \Bigr) ({\bf e}_f \cdot {\bf n}_e)^2 \cdot \mid \frac{1}{p_e} \int_{0}^{+\infty} \psi_i(r) \sin(p_e r) 
 r dr \mid^2 do \nonumber \\
 &=& \alpha a^{2}_{0} \frac{8 \pi p^{3}_e}{\omega} ({\bf e}_f \cdot {\bf n}_e)^2 \cdot \mid \frac{1}{p_e} \int_{0}^{+\infty} \exp(-\gamma r) \sin(p_e r) dr \mid^2 do
 \; \; \; \label{cr-sec1}
\end{eqnarray}
where ${\bf p}_e = p_e {\bf n}_e$ and ${\bf n}_e$ is the unit vector which determines the direction of propagation of the emitted photo-electron. Also, in this equation here $\alpha$ is the fine 
structure constant and $a_0$ is the Bohr radius and $\gamma = t = \sqrt{2 I_1}$. A few steps of additional transformations lead to the following final formula
\begin{eqnarray}
 d\sigma({\rm H}^-) = 8 \pi C^2 \alpha a^{2}_{0} \Bigl( \frac{p_e}{p^{2}_e + \gamma^2} \Bigr)^3 ({\bf k}_f \times {\bf n}_e)^2 do \; \; \; \label{cr-sec2}
\end{eqnarray}
where we used the identity $\overline{({\bf e}_f \cdot {\bf n}_e)^2} = \frac12 ({\bf k}_f \times {\bf n}_e)^2$ and the fact that the energy conservation law (for the photon) is written in the form: 
$\hbar \omega = \frac{p^2_e}{2 m_e} + \frac{\gamma^2}{2}$, or in atomic units: $\omega = \frac{p^2_e}{2} + \frac{\gamma^2}{2}$. To transform the formula from Eq.(\ref{cr-sec}) to its final form we 
have chosen the incident (one-electron!) wave function in the form of Eq.(\ref{eq32}), while the final wave function $\psi_{f}$ is written in the form $exp(\imath {\bf p}_e \cdot {\bf r})$. Also in 
this equation $p_e = \mid {\bf p}_e \mid$ is the momentum of the photo-electron (in atomic units) and $\gamma$ = 0.23558869473885 (see Table I). 

By integrating over angular variables ($do = \sin\theta d\theta d\phi$) one finds the total cross-section of the photodetachment  
\begin{equation}
 \sigma({\rm H}^-) = \frac{64 \pi^2}{3} \alpha a^{2}_{0}\cdot C^2 \cdot \frac{p^3_e}{(p^2_e + \gamma^2)^3} = 33.9954673 \cdot 10^{-18} \cdot C^2 \cdot \frac{p^3_e}{(p^2_e + \gamma^2)^3} 
  \; \; \; cm^2 \; \; \; .
\end{equation}
where $C$ is the constant from Eq.(\ref{eq32}). This formula with the known value of $C$ can directly be used in calculations of the photodetachment cross-section $\sigma(H^-)$. This formula 
essentially coincides with Eq.(8) from \cite{Ohm}. In our case, $C = 0.3562404$, and therefore, from the last formula one finds
\begin{equation}
 \sigma({\rm H}^-) = 4.31427025 \cdot 10^{-18} \cdot \frac{p^{3}_{e}}{(p^{2}_{e} + \gamma^2)^3} \; \; \; cm^2 \; \; \; . \label{simple}
\end{equation}
This simple method predicts the maximum of the photodetachment cross-section $\sigma_{max} \approx$ 41.24335828$\cdot 10^{-18}$ $cm^2$, which is close to the theoretically `exact' maximal value 
\cite{PRA1994}, but slightly above it.   

In actual calculations the overall accuracy of the simple formula, Eq.(\ref{simple}), is outstanding. This remarkable fact follows, in part, from the comparison of the sizes of atomic system $a$ 
with the wavelengths $\lambda$ of the photons which produce photodetachment of the H$^{-}$ ion(s). It is clear that $a \approx a_0 \approx 0.529 \AA$ where $a_0$ is the Bohr radius and $a$ is the 
effective radius of the atomic H$^{-}$ ion. On the other hand, the wavelengths $\lambda$ of the photons which produce the photodetachment is $\lambda \approx 1 \cdot 10^5 \AA$. This means that 
$\lambda \gg a$, or, in other words, the ratio $\frac{\lambda}{a} \approx 2 \cdot 10^{5}$ is a very large number. Briefly, we can say that the incident photon sees only an outer-most asymptotics 
of the three-body wave function. All details of the wave function of the H$^{-}$ ion at small electron-nucleus distances are not important to describe the photodetachment of this ion. 

To conclude this Section we have to note that some important details of the photodetachment of the H$^{-}$ ion cannot be investigated by applying this simple method only. For instance, it 
cannot be used, in principle, to evaluate the transition probabilities for the $ps-$channel(s) and/or predict the final probabilities of the formation of hydrogen atoms in the excited 
$s-$states for the $sp-$channels. Nevertheless, this method is often used in astrophysics for fast evaluation of the absorbtion coefficient of the H$^{-}$ ion at different conditions.

\section{Inverse bremsstrahlung in the field of hydrogen atom}

The knowledge of the photodetachment cross-sections for the $sp-$channels allows one to evaluate the absorbtion of infrared and visible radiation by the H$^{-}$ ion in photospheres of 
many stars with surface temperatures $T_s \le$ 8,250 $K$. Additional contribution from the $ps-$channels increases the overall accuracy of such evaluations. However, at large 
wavelengths ($\lambda \ge 12,000 \AA$) a different mechanism begin to play a significant role in the absorbtion of infrared radiation. Radiation at these wavelengths can be absorbed by 
`free' electrons which move in the fileds of hydrogen atoms. In reality, such an absorbtion by `free' electrons lead to the disappearance of a noticeable amount of propagating infrared 
radiation with $\lambda \ge 12,000 \AA$. It cannot be explained by the photodetachment of the H$^{-}$ ions. Therefore, in this Section we need to discuss the electron scattering 
in the field of the heavy hydrogen atom, Eq.(\ref{ee0}) and radiation transfer which arise during transitions between the `unbound' states of moving electrons. Note that during such 
a process the energy of solar radiation is transferred, in part, to the scattered electrons. Electrons are accelerated, but the mean energy of the radiation field decreases. In classical 
physics this process is called inverse bremsstrahlung. Inverse bremsstrahlung is important in many areas of modern physics and various applications. In stellar photospheres of hydrogen 
stars inverse bremsstrahlung is responsible for absorption of infrared radiation with $\lambda \ge$ 12,000 $\AA$. This includes our Sun \cite{Ohm2}. 

As mentioned above at small energies of the incident photon $h \nu \le 5 eV$ the original (ground) state of the hydrogen atom does not change. This means that for photons of small energies 
we can ignore all possible excitations of the hydrogen atom, i.e. atomic transitions from the ground state of the H atom into other excited states. The incident photon is absorbed 
exclusively by the electron which moves in the field of the neutral hydrogen atom. In this case the electron scattering at the hydrogen atom (H + e$^{-}$ + $h \nu$ = H + e$^{-}({*})$) can 
be considered as a regular (but non-elastic!) scattering of an electron in the field of the heavy hydrogen atom H which is electrically neutral. In the lowest order approximation we can 
replace the actual hydrogen atom by its polarization potential $V(r) = -\frac{a}{r^{4}}$, where $a$ is a positive constant. Note that the potential $V(r)$ is a central and local potential. 
For the ${}^{\infty}$H atom the numerical value of $a$ in atomic units equals $a = \frac94$. This means that we need to consider the electron scattering in the field of the following 
potential
\begin{equation}
   V(r; a, b) = - \frac{a}{(r + b)^{4}} \label{poten}
\end{equation}
which analytically depends upon the two real parameters $a$ and $b$. It is clear that we are dealing with the case of `polarization' inverse bremsstrahlung, which is substantially different 
from the regular (or Coulomb) bremsstrahlung (see, e.g., \cite{Consal}). Our goal is to determine the cross-section of inverse bremsstrahlung for an electron which moves in the field of the 
potential $V(r; a, b)$, Eq.(\ref{poten}). By using different numerical values for the $a$ and $b$ parameters we can obtain a variety of such cross-sections. The main interest, however, is 
related to the cases when in Eq.(\ref{poten}) $a \approx 2.25$ and $b \approx 0$, respectively. 

To calculate the cross-section of inverse bremsstrahlung in the field of the neutral hydrogen atom we can use a number of different methods which can be separated into two large groups. The first 
group of methods is based on the use of very approximate electronic wave functions which do not describe electron-electron correlations. However, by using these wave functions we can solve `exactly' 
the corresponding scattering problem. Moreover, by varying a number of non-linear parameters included in the incident and final wave functions one can obtain a good agreement with the known 
experimental data. The second group of methods is based on an explicit construction of the two-electron wave functions. In this approach all possible electron correlations are included into wave 
functions. For bound states such wave functions can be determined to very high numerical accuracy. However, it is very difficult to obtain accurate numerical solutions of the general scattering 
problem with truly correlated wave functions. The source of the problems is well known, since, e.g., the total number of independent variables is different for the incident and final wave functions. 
 In this study we develop an improved version of single-electron approach. In our approach the hydrogen atom is replaced by an `effective' central potential. This interaction potential explicitly 
depends upon a few parameters which can be varied to achieve a better approximation to actual scattering data. 

Let us describe the method which can be used to calculate cross-section of inverse bremsstrahlung in the field of the hydrogen atom. First, we need to construct the incident and final wave 
functions which are the stationary one-electron wave functions of the continuous spectrum of the radial Schr\"{o}dinger equation (see, e.g., \cite{LLQ}) written for one electron which moves 
in the field of the central potential, Eq.(\ref{poten}). The unknown wave function $\Psi(r)$ can be represented in the form $\Psi_{\ell}(r) = A_{\ell}(r) \exp(\imath \delta_{\ell}(r))$, where 
$A_{\ell}(r)$ is the amplitude, while $\delta_{\ell}(r)$ is the phase of the wave function. Note that in the field of any central potential the angular momentum $\ell$ is a conserving quantum
number (or good quantum number). It is clear that it should be an additional relation/equation between the amplitude $A_{\ell}(r)$ and a phase $\delta_{\ell}(r)$ of the wave function, since we 
cannot simply replace one unknown function by two new functions. By using this `additional' equation one reduces the original problem to the solution of the following differential equation
\begin{equation}
 \frac{d \delta_{\ell}(r)}{dr} = - \frac{2 V(r; a, b)}{k} \Bigl[ j_{\ell}(k r) \cos \delta_{\ell}(r) - n_{\ell}(k r) \sin \delta_{\ell}(r)
 \Bigr]^2 \label{perphas}
\end{equation}
where the phase $\delta_{\ell}(r)$ of the wave function is assumed to be explicitly $r-$dependent. The initial condition for the phase is written in the form $\delta_{\ell}(r = 0)$. 
With this initial condition Eq.(\ref{perphas}) is reduced to the form of the following integral equation 
\begin{equation}
 \delta_{\ell}(r) = - \frac{2}{k} \int_{0}^{r} V(s; a, b) \Bigl[ j_{\ell}(k s) \cos \delta_{\ell}(s) - n_{\ell}(k s) \sin \delta_{\ell}(s)
 \Bigr]^2 ds \label{phas1}
\end{equation}
The equations Eq.(\ref{perphas}) and/or Eq.(\ref{phas1}) are the main equations of the variable phase method. The method of variable phase has been developed by V.V. Babikov and others in the 
middle of 1960's \cite{Babik} (see also references therein). Each of the two equations Eqs.(\ref{perphas}) - (\ref{phas1}) can be solved to high accuracy by using various numerical methods. In 
general, if we know the phase function $\delta_{\ell}(r)$ for each $r$, then it is possible to determine the amplitude $A_{\ell}(r)$ of the wave function and the wave function $\Psi_k(r) = 
A_{\ell}(r) \exp(\imath \delta_{\ell}(r))$ itself. The corresponding equation for the amplitude $A_{\ell}(r)$ takes the form \cite{Babik}
\begin{eqnarray}
 \frac{d A_{\ell}(r)}{dr} = - \frac{A_{\ell}(r) V(r; a, b)}{k} \Bigl[ j_{\ell}(k r) \cos \delta_{\ell}(r) - n_{\ell}(k r) \sin \delta_{\ell}(r)
 \Bigr] \Bigl[ j_{\ell}(k r) \sin \delta_{\ell}(r) \nonumber \\ 
 + n_{\ell}(k r) \cos \delta_{\ell}(r) \Bigr] \label{amplit}
\end{eqnarray}
Note that the variable phase method works very well for central potentials which rapidly vanish at $r 
\rightarrow +\infty$, e.g., for $\mid V(r) \mid \sim r^{-n}$, where $n \ge 3$ at $r \rightarrow +\infty$. 
This means that we can apply this method to determine the accurate wave functions of the continuous spectra 
in the field of the polarization potential, Eq.(\ref{poten}), created by the neutral H atom which is 
assumed to be infinitely heavy. It appears that the variable phase method provides a large number of 
advantages in applications to the problems related with the absorption/emission of radiation from the states 
of continuous spectra.

In the case of inverse bremsstrahlung the energy conservation law is written in the form 
\begin{equation}
  \frac{\hbar k^{2}_{i}}{2 m_e} + \omega = \frac{\hbar k^{2}_{f}}{2 m_e} \; \; \; , \; \; or \; \; \;  \hbar \omega = 
  \frac{p^{2}_{f}}{2 m_e} - \frac{p^{2}_{i}}{2 m_e} \label{energ}
\end{equation}
where $k_i$ and $k_f$ are the wave numbers defined as $k_a = \mid {\bf k}_a \mid = \frac{\mid {\bf p}_a \mid}{\hbar}$ and ${\bf p}_a$ is the 
momentum of the particle $a$ and ${\bf k}_a$ is the corresponding wave vector. In atomic units ${\bf k}_a = {\bf p}_a$ and $k_a = p_a$, while
Eq.(\ref{energ}) takes the form $\omega = \frac12 (p^{2}_{f} - p^{2}_{i})$. Also, from Eq.(\ref{energ}) one finds $k_f = \sqrt{k^{2}_{i} + 
\frac{2 m_e}{\hbar} \omega}$. By using the variable phase method described above we can obtain the initial and final state wave functions, i.e. 
$\Psi_{k_i}(r)$ and $\Psi_{k_f}(r)$, respectively. In the variable phase method these wave functions are represented as the products of their 
amplitudes and phase parts, i.e. $\Psi_{\ell;k_i}(r) = A_{\ell;k_i}(r) \exp(\imath \delta_{\ell;k_i}(r))$ and $\Psi_{\ell;k_f}(r) = 
B_{\ell;k_f}(r) \exp(\imath g_{\ell;k_f}(r))$. Let us assume that by using the procedure described above we have determined the incident and 
final wave functions. Now, these two wave functions can be used in the following computations of the cross-section of inverse bremsstrahlung. 

The explicit formula for the cross-section of inverse bremsstrahlung is (in relativistic units)
\begin{equation}
  d\sigma_{{\bf k}{\bf p}_f} = \frac{\omega e^2}{(2 \pi)^4 m_e p_i} \mid {\bf e} \cdot {\bf p}_{i \rightarrow f} \mid^2 do_{{\bf k}} d^3{\bf p}_f  
 \label{formula}
\end{equation}
where $p_i = \mid {\bf p}_i \mid$ is the `absolute' momentum of the incident electron, ${\bf p}_f$ is the momentum of the final electron and 
${\bf p}_{i \rightarrow f}$ is the following matrix element
\begin{equation}
  {\bf p}_{i \rightarrow f} = - \imath \int \Psi^{*}_{k_f} \nabla \Psi_{k_i} d^3{\bf r} \label{moment}
\end{equation}
where $\Psi_{k_i}$ and $\Psi_{k_f}$ are the incident and final wave functions. By integrating Eq.(\ref{formula}) over all directions of the final momentum one 
finds the following formula for the cross-section (in atomic units):
\begin{equation}
  d\sigma_{{\bf k} p_f} = \alpha^3 a^{2}_{0} \frac{p^{2}_{f} - p^{2}_{i}}{(2 \pi)^3 p_i} \mid {\bf e} \cdot {\bf p}_{i \rightarrow f} \mid^2 
  do_{{\bf k}} p^2_f d p_f \label{form1}
\end{equation}
After a few steps of straightforward transformations one finds from this formula
\begin{equation}
  d\sigma_{p_i p_f} = \frac{2 \alpha^3 a^{2}_{0}}{3 \pi} \frac{(p^{2}_{f} - p^{2}_{i}) p^2_f}{p_i} \mid \int^{\infty}_0 \Psi_f 
  \Bigl(\frac{d \Psi_{i}}{dr} \Bigr) dr \mid^2 d p_f \label{formula2}
\end{equation}
By using Eqs.(\ref{perphas}) and (\ref{amplit}) we obtain the following expression for the derivative of the $\Psi_i$ function:
\begin{eqnarray}
 \frac{d \Psi_{i}}{dr} &=& \frac{d A_{\ell;k_i}(r)}{dr} \exp(\imath \delta_{\ell;k_i}) +  \imath A_{\ell;k_i}(r) \exp(\imath \delta_{\ell;k_i})
 \frac{d \delta_{\ell;k_i}(r)}{dr} \label{eqphase} \\
 &=& - \frac{V(r; a, b) A_{\ell;k}(r) \exp(\imath \delta_{\ell;k_i}) }{k} \Bigl[ j_{\ell}(k r) \cos \delta_{\ell}(r) - n_{\ell}(k r) \sin \delta_{\ell}(r)
 \Bigr] \Bigl[ j_{\ell}(k r) \sin \delta_{\ell}(r)  \nonumber \\
 &+& n_{\ell}(k r) \cos \delta_{\ell}(r) \Bigr] - \frac{2 V(r; a, b)}{k} A_{\ell;k_i}(r) \exp(\imath \delta_{\ell;k_i}) \Bigl[ j_{\ell}(k r) \cos 
 \delta_{\ell}(r) - n_{\ell}(k r) \sin \delta_{\ell}(r) \Bigr]^2 \nonumber
\end{eqnarray}
or, in other words,
\begin{eqnarray}
 \frac{d \Psi_{i}}{dr} &=& - \frac{V(r; a, b)}{k} \Psi_{i} \Bigl[ j_{\ell}(k r) \cos \delta_{\ell}(r) - n_{\ell}(k r) \sin \delta_{\ell}(r)
 \Bigr] \Bigl[ j_{\ell}(k r) \sin \delta_{\ell}(r) + n_{\ell}(k r) \cos \delta_{\ell}(r) \Bigr] \nonumber \\
 &-& \frac{2 V(r; a, b)}{k} \Psi_{i} \Bigl[ j_{\ell}(k r) \cos \delta_{\ell}(r) - n_{\ell}(k r) \sin \delta_{\ell}(r) \Bigr]^2 \nonumber
\end{eqnarray}
The last equation does not include explicitly the amplitude function $A_{\ell;k_i}(r)$. Note also that the last equation can be re-written in a number of different forms which are 
more convenient in some cases. 

Another simplification follows from the fact that the wavelength $\lambda$ of the light quantum which produces photodetachment of the H$^{-}$ ion is $\ge$ 12,000 $\AA$, while the 
effective `scattering zone' $R_s$ for the process, Eq.(\ref{ee0}) is less than 15 - 25 $\AA$. Therefore, we can apply the approximation based on small values of the dimensionless 
parameter $\frac{R_s}{\lambda} \ll 1$. This approximation allows one to replace the $r-$dependent phase factor $\delta_{\ell;k_i}(r)$ in the wave function by its asymptotic value 
$\delta_{\ell;k_i}(+\infty)$ determined at the infinity. In other words, the phase of the wave function does not depend upon $r$, i.e. it is a constant. The wave function 
$\Psi_{\ell;k_i}(r)$ is represented in the form $\Psi_{\ell;k_i}(r) = A_{\ell;k_i}(r) \exp(\imath \delta_{\ell;k_i})$, where $\delta_{\ell} = \delta_{\ell}(r = +\infty)$ is a 
constant, or almost a constant. This drastically simplifies numerical computations of all integrals needed in Eq.(\ref{formula2}). Finally, it should be noted that based on the 
approach developed here we can evaluate the total cross-section of inverse bremsstrahlung for electrons scattered by the neutral hydrogen atoms in stellar photospheres. 

\section{Conclusion}

We have considered the absorption of infrared and visible radiation by the negatively charged hydrogen ions H$^{-}$. Such an absorption of radiation plays an extremely important 
role in Solar and stellar astrophysics. In general, absorption of infrared and visible radiation by the H$^{-}$ ions is mainly related with the photodetachment of these ions. The 
absorption of radiation by electrons scattered in the field of the neutral hydrogen atom(s) also contributes to the radiation absorbtion coefficient at very large wavelength 
$\lambda \ge 12,000 \AA$ (infrared radiation). In this study we discuss all details important in calculations of the photodetachment cross-section with the use of highly accurate 
(or truly correlated) wave functions of the two-electron (or three-body) H$^{-}$ ion. Photodetachment cross-sections of the ${}^{\infty}$H$^{-}$, ${}^{3}$H$^{-}$ (or T$^{-}$), 
${}^{2}$H$^{-}$ (or D$^{-}$) and ${}^{1}$H$^{-}$ ions are determined to high accuracy and for a large number of photo-electron momenta/energies. Some additional details of our methods 
and computational results will be discussed in our next study \cite{Fro2014}. 

Right now, we can only note that our current calculations performed for the photodetachment cross-section for the ${}^{\infty}$H$^{-}$ ion indicate a good agreement with the results 
obtained for this ion by other methods (see, e.g., \cite{PRA2014} - \cite{PRA1994}). For the negatively charged hydrogen ions with the finite nuclear masses calculations the 
photodetachment cross-sections with highly accurate wave functions have never been performed. In this sense our results for the ${}^{3}$H$^{-}$ (or T$^{-}$), ${}^{2}$H$^{-}$ (or D$^{-}$) 
and ${}^{1}$H$^{-}$ ions from Tables VI and VII are unique. Furthermore, this paper opens a new avenue for accurate computations of the photodetachment cross-section of all negatively 
charged hydrogen ions with the use of highly accurate wave functions. In future calculations there is no need to reduce the overall accuracy of the truly correlated wave functions by 
using different approximate procedures, e.g., by representing these wave functions in the form of natural orbital expansion(s). Currently, we are working on the problem of incorporating 
extremely accurate wave functions with 3500 - 4000 basis functions (see Table I) in calculations of the photodetachment cross-sections.  

\begin{center}
  {\Large Appendix}
\end{center}

In this Appendix we test our variational wave functions constructed for the negatively charged hydrogen ions (${}^{\infty}$H$^{-}, {}^{3}$H$^{-}, {}^{2}$H$^{-}$ 
and ${}^{1}$H$^{-}$) in actual computations of the lowest order QED corrections for these two-electron systems. The lowest order QED corrections for 
the ground $1^1S-$state in the negatively charged hydrogen ${}^{\infty}$H$^{-}$ ion is determined by the formula (in atomic units) 
\begin{eqnarray}
 \Delta E^{QED}_{\infty} &=& \frac{8}{3} Q \alpha^3 \Bigl[ \frac{19}{30} - 2 \ln \alpha - \ln K_0 \Bigr] \langle \delta({\bf r}_{eN}) \rangle 
   \nonumber \\
 &+& \alpha^3 \Bigl[ \frac{164}{15} + \frac{14}{3} \ln \alpha - \frac{10}{3} S(S + 1) \Bigr] \langle \delta({\bf r}_{ee}) \rangle - \frac{7}{6 \pi} \alpha^3 
     \langle \frac{1}{r^{3}_{ee}} \rangle \label{LQED}
\end{eqnarray}
where $\alpha$ is the fine structure constant, $Q$ is the nuclear charge (in atomic units) and $S$ is the total electron spin. For the ground states in 
all H$^{-}$ ions considered in this study we have $Q = 1$ and $S = 0$ (singlet states) in Eq.(\ref{LQED}). The notation $\ln K_0$ in this formula stands for the Bethe 
logarithm \cite{BS}. To simplify numerical calculations in this Appendix one can take the value of $\ln K_0$ from Table 3 in \cite{Fro2005}. 

For the two-electron ions with finite nuclear mass we need to evaluate the corresponding recoil correction to the lowest-order QED correction $\Delta E^{QED}_{\infty}$,
Eq.(\ref{LQED}). Such a correction is calculated with the use of the following formula (in atomic units) 
\begin{eqnarray}
 \Delta E^{QED}_{M} &=& \Delta E^{QED}_{\infty} - \Bigl(\frac{2}{M} + \frac{1}{M + 1} \Bigr) \Delta E^{QED}_{\infty} + \frac{4 \alpha^3 Q^2}{3 M} 
 \Bigl[ \frac{37}{3} - \ln \alpha - 4 \ln K_0 \Bigr] \langle \delta({\bf r}_{eN}) \rangle \nonumber \\
 &+& \frac{7 \alpha^3}{3 \pi M} \langle \frac{1}{r^{3}_{eN}} \rangle \label{LQEDA}
\end{eqnarray} 
where $M \gg m_e$ is the nuclear mass. All expectation values in this equation must be determined for the real two-electron ions, i.e. ions with the finite nuclear 
masses. The inverse mass $\frac{1}{M}$ is a small dimensionless parameter which for the ions considered in this study  is smaller than $\le 5 \cdot 10^{-4}$. The last 
terms in both Eq.(\ref{LQED}) and Eq.(\ref{LQEDA}) are usually called the Araki-Sucher terms, or Araki-Sucher corrections \cite{Araki}, \cite{Sucher}, since this 
correction was obtained and investigated for the first time in papers by Araki and Sucher. The expectation value of the term $\langle \frac{1}{r^{3}_{ee}} 
\rangle$ is singular, i.e. it contains the regular and non-zero divergent parts. General theory of the singular three-body integrals was developed in our earlier works 
(see, e.g., \cite{Fro2005} and references therein). In particular, in \cite{Fro2005} we have shown that the $\langle \frac{1}{r^{3}_{ee}} \rangle$ expectation value is 
determined by the formula
\begin{eqnarray}
  \langle \frac{1}{r^{3}_{ee}} \rangle = \langle \frac{1}{r^{3}_{ee}} \rangle_R + 4 \pi \langle \delta({\bf r}_{ee}) \rangle
\end{eqnarray} 
where $\langle \frac{1}{r^{3}_{ee}} \rangle_R$ is the regular part of this expectation value and $\langle \delta({\bf r}_{ee}) \rangle$ is the expectation 
value of the electron-electron delta-function. Briefly, we can say that the overall contribution of the signular part of the $\frac{1}{r^{3}_{ee}}$ operator 
is reduced to the expectation value of the corresponding delta-function.

It follows from the formulas Eqs.(\ref{LQED}) and (\ref{LQEDA}) that highly accurate evaluation of the lowest-order QED correction is a very effective test
of the non-relativistic wave functions of the negatively charged hydrogen ions. After this test the bound state wave functions can be used in numerical calculations of 
the photodetachment cross-section(s) and absorption coefficients of the negatively charged hydrogen ions. Indeed, the formulas Eqs.(\ref{LQED}) and (\ref{LQEDA}) 
contain both the electron-nuclear and electron-electron delta-functions. To prove the correctness of these values and approximately evaluate the total number of stable 
decimal digits in each of the delta-functions one need to compute the corresponding cusp values and compare them with the expected (or predicted) values. This essentially 
coincides with the standard test for the non-relativistic wave functions. In addition to these delta-functions Eqs.(\ref{LQED}) and (\ref{LQEDA}) include the 
$\langle \frac{1}{r^{3}_{ij}} \rangle$ expectation value which is singular. Accurate numerical evaluation of the $\langle \frac{1}{r^{3}_{eN}} \rangle$ and
$\langle \frac{1}{r^{3}_{ee}} \rangle$ expectation values is not an easy task (see, e.g., \cite{Fro2005}), but it provides another effective test for the 
non-relativistic wave function. Our results for the lowest-order QED correction for the ${}^{\infty}$H$^{-}, {}^{3}$H$^{-}, {}^{2}$H$^{-}$ and ${}^{1}$H$^{-}$ 
ions can be found in Table V (in atomic units). The lowest-order QED corrections $\Delta E^{QED}_{\infty}$ and $\Delta E^{QED}_{M}$ determined for each of these ions 
with the finite nuclear masses are the most accurate values to-date. In reality, they coincide in seven first decimal digits with the corresponding values from 
\cite{Fro2005}, where we used less accurate variational wave functions. Very likely, the lowest-order QED corrections are of interest by themselves, but we 
have calculated them here only to prove very high quality of our non-relativistic wave functions.

\newpage
\begin{table}[tbp]
   \caption{The non-relativistic total energies $E$ (in $a.u.$) determined with different number 
            of basis functions $N$, the asymptotic values of the total energies $E(N = \infty)$ 
            (in $a.u.$) and ionization/detachment potential $\chi_1$ (in $eV$) of the hydrogen 
            negatively charged ions (isotopes) in atomic units.}
     \begin{center}
     \begin{tabular}{| c | c | c | c | c |}
      \hline\hline
 $K$ & $E$(${}^{\infty}$H$^{-}$) & & &  $E$(${}^{3}$H$^{-}$) \\
     \hline
 3500 & -0.527751 016544 377196 590213 & & & -0.527649 048201 920733 538766 \\
 
 3700 & -0.527751 016544 377196 590333 & & & -0.527649 048201 920733 538885 \\

 3840 & -0.527751 016544 377196 590389 & & & -0.527649 048201 920733 538983 \\

 4000 & -0.527751 016544 377196 590446 & & & -0.527649 048201 920733 539050 \\
     \hline
 $E(N = \infty)$ & -0.527751 016544 377196 59075(10) & & & -0.527649 048201 920733 539355(10) \\
     \hline
 $\chi_1$ & -0.755143 903366 701124 238 & & & -0.754843 900893 517142 964 \\
        \hline\hline

 $K$ & $E$(${}^{2}$H$^{-}$) & & &  $E$(${}^{1}$H$^{-}$) \\

        \hline
 3500 & -0.527598 324689 706528 595827 & & & -0.527445 881119 767477 071142 \\

 3700 & -0.527598 324689 706528 595960 & & & -0.527445 881119 767477 071261 \\

 3840 & -0.527598 324689 706528 596044 & & & -0.527445 881119 767477 071359 \\

 4000 & -0.527598 324689 706528 596110 & & & -0.527445 881119 767477 071425 \\
     \hline
$E(N = \infty)$ & -0.527598 324689 706528 596355(10) & & & -0.527445 881119 767477 071665(10) \\
     \hline
  $\chi_1$ & -0.754694 721951 430623 253 & & & -0.754246 603605 793792 368 \\
    \hline \hline
  \end{tabular}
  \end{center}
  \end{table}
%


\begin{table}[tbp]
   \caption{Convergence of the expectation values of some electron-nuclear and electron-electron properties for the 
    ${}^{\infty}$H$^{-}$ ion. All values are shown in atomic units and $N$ is the total number of basis functions used.}
     \begin{center}
     \begin{tabular}{| c | c | c | c |}
      \hline\hline
  & $\langle r_{eN} \rangle$ & $\langle r^2_{ee} \rangle$ &  $\langle r^{-2}_{eN} \rangle$ \\ 
     \hline
 3500 & 2.710178278444420365208 & 25.20202529124033189252 & 1.11666282452545228 \\ 
 
 3700 & 2.710178278444420365265 & 25.20202529124033189563 & 1.11666282452544953 \\ 

 3840 & 2.710178278444420365286 & 25.20202529124033189660 & 1.11666282452544421 \\ 

 4000 & 2.710178278444420365301 & 25.20202529124033189738 & 1.11666282452543860 \\ 
        \hline
 & $\tau_{eN}^{(a)}$ & $\langle -\frac12 \nabla^2_e \rangle$ & $\langle {\bf r}_{eN} \cdot {\bf r}_{ee} \rangle$ \\
       \hline
 3500 & 0.6451445424122193894344 & 0.26387550827218859829282 & 12.60101264562016594726 \\

 3700 & 0.6498715811920881669355 & 0.26387550827218859829304 & 12.60101264562016594781 \\

 3840 & 0.6498715811920881669351 & 0.26387550827218859829337 & 12.60101264562016594830 \\

 4000 & 0.6498715811920881669349 & 0.26387550827218859829358 & 12.60101264562016594869 \\
     \hline \hline
  \end{tabular}
  \end{center}
${}^{(a)}$The notation $\tau_{eN}$ stands for the $\langle \cos ({\bf r}_{31} {}^{\wedge} {\bf r}_{21}) \rangle = 
\langle \cos ({\bf r}_{eN} {}^{\wedge} {\bf r}_{ee}) \rangle$ expectation value. Note also, that for all two-electron atomic systems 
$2 \tau_{eN} + \tau_{ee} = 1 + 4 \langle f \rangle$, where $\langle f \rangle$ is the expectation value mentioned in Table III.
  \end{table}


\begin{table}[tbp]
   \caption{The expectation values of some propeties (in atomic units) for the ${}^{\infty}$H$^{-}$ ion.}
     \begin{center}
     \begin{tabular}{| c | c | c | c |}
      \hline\hline
 $\langle r^{-2}_{eN} \rangle$ & $\langle r^{-2}_{ee} \rangle$ & $\langle r^{-1}_{eN} \rangle$ & $\langle r^{-1}_{ee} \rangle$ \\
      \hline 
 1.11666282452542572 & 0.15510415256242466 & 0.6832617676515272224 & 0.311021502214300052 \\
       \hline
 $\langle r_{eN} \rangle$ & $\langle r_{ee} \rangle$ & $\langle r^{2}_{eN} \rangle$ & $\langle r^{2}_{ee} \rangle$ \\
      \hline
 2.7101782784444203653 & 4.4126944979917277211 & 11.913699678051262274 & 25.202025291240331897 \\
      \hline
 $\langle r^{3}_{eN} \rangle$ & $\langle r^{3}_{ee} \rangle$ & $\langle r^{4}_{eN} \rangle$ & $\langle r^{4}_{ee} \rangle$ \\
      \hline
 76.02309704902717911 & 180.60560023017477483 & 645.144542412219375 & 1590.09460393948530 \\
      \hline\hline
 $\langle [r_{32} r_{31}]^{-1} \rangle$ & $\langle [r_{eN} r_{ee}]^{-1} \rangle$ & $\langle [r_{32} r_{31} r_{21}]^{-1} \rangle$ & $\langle \delta({\bf r}_{eeN}) \rangle$ \\
     \hline
 0.38262789034020545 & 0.25307756706456687 & 0.20082343962918944 & 5.129778775490$\cdot 10^{-3}$ \\
     \hline\hline
 $\langle \delta({\bf r}_{eN}) \rangle$ & $\nu_{eN}^{(a)}$ & $\langle \delta({\bf r}_{eN}) \rangle$ & $\nu_{ee}^{(a)}$ \\
      \hline
 0.1645528728473590 & -1.00000000001778 & 2.737992126104611$\cdot 10^{-3}$ & 0.500000002446 \\
     \hline
 $\tau_{eN}$ & $\tau_{ee}$ & $\langle f \rangle$ & $\langle {\bf r}_{31} \cdot {\bf r}_{32} / r^3_{31} \rangle$ \\
      \hline
 0.6498715811920881669 & -0.1051476935659779011 & 0.048648867204549608205 & -0.4642618530806317014 \\
     \hline\hline 
 $\langle -\frac12 \nabla^2_e \rangle$ & $\langle -\frac12 \nabla^2_N \rangle$ & $\langle \nabla_e \cdot \nabla_e \rangle$ & $\langle \nabla_e \cdot \nabla_N \rangle$ \\
      \hline
 0.2638755082721885983 & 0.560630798396681918 & 0.0328797818523047217 & -5.60630798396681918 \\
     \hline
 $\langle {\bf r}_{eN} \cdot {\bf r}_{ee} \rangle$ & $\langle (r^{-3}_{eN})_R \rangle$ & $\langle (r^{-3}_{ee})_R \rangle$ & $\langle r^{-3}_{eN} \rangle$ \\
      \hline
 12.601012645620165948 & -3.43559485054432 & 0.064307887285283 & -1.36776246468689 \\ 
    \hline \hline
  \end{tabular}
  \end{center}
 ${}^{(a)}$The expected cusp values (in $a.u.$) for the ${}^{\infty}$H$^{-}$ ion are $\nu_{eN} = -1.0$ and $\nu_{ee} = 0.5$ (exactly).
  \end{table}
%

\begin{table}[tbp]
   \caption{Convergence of the expectation values of the electron-nuclear and electron-electron delta-functions 
    ($\langle \delta({\bf r}_{eN}) \rangle$ and $\langle \delta({\bf r}_{ee}) \rangle$) and regular parts od the inverse cubic  
    expectation values ($\langle r^{-3}_{eN} \rangle_R$ and $\langle r^{-3}_{ee} \rangle_R$) for the ${}^{3}$H$^{-}, {}^{2}$H$^{-}$ 
    and ${}^{1}$H$^{-}$ ions. All values are shown in atomic units and $N$ is the total number of basis functions used.}
     \begin{center}
     \begin{tabular}{| c | c | c | c |}
      \hline\hline
  & ${}^{3}$H$^{-}$ (tritium) & ${}^{2}$H$^{-}$ (deuterium) & ${}^{1}$H$^{-}$ (protium) \\
     \hline
  & $\langle \delta({\bf r}_{eN}) \rangle$ & $\langle \delta({\bf r}_{eN}) \rangle$ & $\langle \delta({\bf r}_{eN}) \rangle$ \\
     \hline
 3500 & 0.16446163681098 & 0.16441626335994 & 0.16427994412837 \\
 
 3700 & 0.16446163681128 & 0.16441626336024 & 0.16427994412867 \\

 3840 & 0.16446163681069 & 0.16441626335966 & 0.16427994412847 \\

 4000 & 0.16446163681105 & 0.16441626336002 & 0.16427994412844 \\
     \hline
  & $\langle \delta({\bf r}_{ee}) \rangle$ & $\langle \delta({\bf r}_{ee}) \rangle$ & $\langle \delta({\bf r}_{ee}) \rangle$ \\
     \hline
 3500 & 2.735845627669$\cdot 10^{-3}$ & 2.734778360513$\cdot 10^{-3}$ & 2.731572792726$\cdot 10^{-3}$ \\
 
 3700 & 2.735845627652$\cdot 10^{-3}$ & 2.734778360496$\cdot 10^{-3}$ & 2.731572792710$\cdot 10^{-3}$ \\ 

 3840 & 2.735845627712$\cdot 10^{-3}$ & 2.734778360555$\cdot 10^{-3}$ & 2.731572792769$\cdot 10^{-3}$ \\ 

 4000 & 2.735845627776$\cdot 10^{-3}$ & 2.734778360619$\cdot 10^{-3}$ & 2.731572792833$\cdot 10^{-3}$ \\
     \hline
     \hline
  & $\langle r^{-3}_{eN} \rangle_R$ & $\langle r^{-3}_{eN} \rangle_R$ & $\langle r^{-3}_{eN} \rangle_R$ \\
     \hline
 3500 & -3.43332137343396 & -3.43219080091021119 & -3.4287944147910614 \\
 
 3700 & -3.43332137346264 & -3.43219080093883454 & -3.4287944148194641 \\

 3840 & -3.43332137341561 & -3.43219080089185737 & -3.4287944147726374 \\

 4000 &  -3.4333213734460 & -3.43219080092215666 & -3.4287944148027678 \\
     \hline
  & $\langle r^{-3}_{ee} \rangle_R$ & $\langle r^{-3}_{ee} \rangle_R$ & $\langle r^{-3}_{ee} \rangle_R$ \\
     \hline
 3500 & 6.427894892817$\cdot 10^{-2}$ & 6.426455565395$\cdot 10^{-2}$ & 6.422130608689$\cdot 10^{-2}$ \\
 
 3700 & 6.427894893018$\cdot 10^{-2}$ & 6.426455565595$\cdot 10^{-2}$ & 6.422130608885$\cdot 10^{-2}$ \\ 

 3840 & 6.427894892413$\cdot 10^{-2}$ & 6.426455564991$\cdot 10^{-2}$ & 6.422130608281$\cdot 10^{-2}$ \\

 4000 & 6.427894892942$\cdot 10^{-2}$ & 6.426455564520$\cdot 10^{-2}$ & 6.422130608203$\cdot 10^{-2}$ \\
    \hline \hline
  \end{tabular}
  \end{center}
  \end{table}


\begin{table}[tbp]
   \caption{The Bethe logarithm (in $a.u.$) and lowest order QED corrections $E^{QED}$ and $E^{QED}_M$ (in $MHz$) for the ground 
            $1^1S-$state(s) in the ${}^{\infty}$H$^{-}, {}^{3}$H$^{-}, {}^{2}$H$^{-}$ and ${}^{1}$H$^{-}$ ions.}
     \begin{center}
     \begin{tabular}{| c | c | c | c | c |}
      \hline\hline
                    & ${}^{\infty}$H$^{-}$ & ${}^{3}$H$^{-}$ (tritium) & ${}^{2}$H$^{-}$ (deuterium) & ${}^{1}$H$^{-}$ (protium) \\
       \hline 
 $\ln K_0$ ($a.u.$) & 2.993004415 & 2.993011414 & 2.993014897 & 2.993025369 \\
    \hline
 $\Delta E^{QED}$ & 8215.203465 & 8210.661159 & 8208.402174 & 8201.615291 \\
    \hline
 $\Delta E^{QED}_{M}$ & 8215.203465 & 8207.191206 & 8203.207102 & 8191.239514 \\
   \hline \hline
  \end{tabular}
  \end{center}
  \end{table}
\newpage
\begin{table}[tbp]
   \caption{Photodetachment cross-section (in $cm^{2}$) of the negatively charged hydrogen ion in their ground $1^1S-$states. 
           $N$ is the total number of basis function in the trial wave function and $E$ is the total energy (in $a.u.$) for this wave 
           function. The final hydrogen atom is formed in the ground $1s$-state. The notation $E_e$ stands for the kinetic energy of 
           the emitted photo-electron (in $a.u.$).}
     \begin{center}
     \scalebox{0.80}{%
     \begin{tabular}{| c | c | c | c | c | c |}
      \hline\hline
        & ${}^{\infty}$H$^{-}$ &  ${}^{\infty}$H$^{-}$ & ${}^{3}$H$^{-}$ & ${}^{2}$H$^{-}$ & ${}^{1}$H$^{-}$ \\
           \hline
  $E_{{\rm H}^{-}}$  & -0.5277751016544153 & -0.5277751016544252 & -0.527649048201671 & -0.527598324689580 & -0.527445881119594 \\ 
           \hline
  $E_e$  & $\sigma(E_e) (N$ = 300) & $\sigma(E_e) (N$ = 350) & $\sigma(E_e) (N$ = 350) & $\sigma(E_e) (N$ = 350) & $\sigma(E_e) (N$ = 350) \\
          \hline     
 0.0000125 & 0.2959861281153E-17 & 0.2959854603169E-17 &  0.2959608363575E-17 &  0.2959527360520E-17 &  0.2959202302718E-17 \\
 0.0001125 & 0.8795361962978E-17 & 0.8795347268968E-17 &  0.8794584542552E-17 &  0.8794308561005E-17 &  0.8793268085947E-17 \\
 0.0003125 & 0.1438357679058E-16 & 0.1438356148152E-16 &  0.1438220490456E-16 &  0.1438164996520E-16 &  0.1437971473501E-16 \\
 0.0006125 & 0.1957704797715E-16 & 0.1957703477352E-16 &  0.1957494491642E-16 &  0.1957401257879E-16 &  0.1957094252745E-16 \\
 0.0010125 & 0.2425248779858E-16 & 0.2425247518253E-16 &  0.2424946573180E-16 &  0.2424806431545E-16 &  0.2424359742878E-16 \\
 0.0015125 & 0.2831673151310E-16 & 0.2831671287514E-16 &  0.2831258986430E-16 &  0.2831063012238E-16 &  0.2830450229469E-16 \\
 0.0021125 & 0.3170965737411E-16 & 0.3170962354548E-16 &  0.3170421932053E-16 &  0.3170161206699E-16 &  0.3169358006790E-16 \\
 0.0028125 & 0.3440414647798E-16 & 0.3440409146283E-16 &  0.3439728216636E-16 &  0.3439394843234E-16 &  0.3438381596859E-16 \\
 0.0036125 & 0.3640350418486E-16 & 0.3640342883386E-16 &  0.3639514145850E-16 &  0.3639102647578E-16 &  0.3637866660179E-16 \\
 0.0045125 & 0.3773699512169E-16 & 0.3773690666895E-16 &  0.3772712159963E-16 &  0.3772220418799E-16 &  0.3770757332968E-16 \\
            \hline
 0.0055125 & 0.3845424767129E-16 & 0.3845415662293E-16 &  0.3844290725784E-16 &  0.3843720233167E-16 &  0.3842034419486E-16 \\
 0.0066125 & 0.3861926168889E-16 & 0.3861917835696E-16 &  0.3860654758202E-16 &  0.3860010314572E-16 &  0.3858114331217E-16 \\
 0.0078125 & 0.3830463436882E-16 & 0.3830456649096E-16 &  0.3829068002096E-16 &  0.3828357090908E-16 &  0.3826270448851E-16 \\
 0.0091125 & 0.3758644736645E-16 & 0.3758639917081E-16 &  0.3757141647769E-16 &  0.3756373675088E-16 &  0.3754121197964E-16 \\
 0.0105125 & 0.3654007536656E-16 & 0.3654004777150E-16 &  0.3652415173423E-16 &  0.3651600703972E-16 &  0.3649210748110E-16 \\
 0.0120125 & 0.3523701334398E-16 & 0.3523700477057E-16 &  0.3522039120618E-16 &  0.3521189189400E-16 &  0.3518691928177E-16 \\
 0.0136125 & 0.3374269418098E-16 & 0.3374270152781E-16 &  0.3372556954001E-16 &  0.3371682494788E-16 &  0.3369108401747E-16 \\
 0.0153125 & 0.3211518525441E-16 & 0.3211520475640E-16 &  0.3209774852243E-16 &  0.3208886257917E-16 &  0.3206264861165E-16 \\
 0.0171125 & 0.3040460861272E-16 & 0.3040463646812E-16 &  0.3038703877562E-16 &  0.3037810689795E-16 &  0.3035169622355E-16 \\
 0.0190125 & 0.2865311628737E-16 & 0.2865314904517E-16 &  0.2863557664068E-16 &  0.2862668378989E-16 &  0.2860032716430E-16 \\
              \hline   
 0.0210125 & 0.2689526056870E-16 & 0.2689529532263E-16 &  0.2687789595522E-16 &  0.2686911566400E-16 &  0.2684303422298E-16 \\
 0.0231125 & 0.2515861990078E-16 & 0.2515865435015E-16 &  0.2514155528590E-16 &  0.2513294944699E-16 &  0.2510733283533E-16 \\
 0.0253125 & 0.2346456740216E-16 & 0.2346459983308E-16 &  0.2344790756524E-16 &  0.2343952678096E-16 &  0.2341453301346E-16 \\
 0.0276125 & 0.2182909599163E-16 & 0.2182912521773E-16 &  0.2181292608860E-16 &  0.2180481040352E-16 &  0.2178056702623E-16 \\
 0.0300125 & 0.2026363869442E-16 & 0.2026366398150E-16 &  0.2024802545112E-16 &  0.2024020531988E-16 &  0.2021681148092E-16 \\
 0.0325125 & 0.1877584342631E-16 & 0.1877586441441E-16 &  0.1876083674625E-16 &  0.1875333412069E-16 &  0.1873086321621E-16 \\
 0.0351125 & 0.1737027795553E-16 & 0.1737029458465E-16 &  0.1735591275960E-16 &  0.1734874220901E-16 &  0.1732724483298E-16 \\
 0.0378125 & 0.1604905302160E-16 & 0.1604906546358E-16 &  0.1603535117581E-16 &  0.1602852097992E-16 &  0.1600802796753E-16 \\
 0.0406125 & 0.1481236028999E-16 & 0.1481236888700E-16 &  0.1479933251391E-16 &  0.1479284568916E-16 &  0.1477337109967E-16 \\
 0.0435125 & 0.1365892760874E-16 & 0.1365893281871E-16 &  0.1364657529017E-16 &  0.1364043052627E-16 &  0.1362197445888E-16 \\
       \hline\hline
  \end{tabular}}
  \end{center}
  \end{table}
\newpage
\begin{table}[tbp]
   \caption{Photodetachment cross-section (in $cm^{2}$) of the negatively charged hydrogen ions in their ground $1^1S-$states
            for intermediate (kinetic) energies of photo-electron $E_e$. All notations are the same as in the previous Table.}
     \begin{center}
     \scalebox{0.80}{%
     \begin{tabular}{| c | c | c | c | c | c |}
      \hline\hline
        & ${}^{\infty}$H$^{-}$ &  ${}^{\infty}$H$^{-}$ & ${}^{3}$H$^{-}$ & ${}^{2}$H$^{-}$ & ${}^{1}$H$^{-}$ \\
           \hline
  $E_e$  & $\sigma(E_e) (N$ = 300) & $\sigma(E_e) (N$ = 350) & $\sigma(E_e) (N$ = 350) & $\sigma(E_e) (N$ = 350) & $\sigma(E_e) (N$ = 350) \\
          \hline    
 0.0465125 & 0.1258639757846E-16 & 0.1258639992733E-16 &  0.1257471444683E-16 &  0.1256890694264E-16 &  0.1255145811860E-16 \\
 0.0496125 & 0.1159163734473E-16 & 0.1159163738578E-16 &  0.1158061097734E-16 &  0.1157513317341E-16 &  0.1155867124808E-16 \\
 0.0528125 & 0.1067098826764E-16 & 0.1067098654801E-16 &  0.1066060141832E-16 &  0.1065544362939E-16 &  0.1063994122132E-16 \\
 0.0561125 & 0.9820464104618E-17 & 0.9820461139031E-17 &  0.9810695855883E-17 &  0.9805846809755E-17 &  0.9791271250301E-17 \\
 0.0595125 & 0.9035905851910E-17 & 0.9035902103435E-17 &  0.9026732599035E-17 &  0.9022179890367E-17 &  0.9008494708252E-17 \\
 0.0630125 & 0.8313100637491E-17 & 0.8313096504894E-17 &  0.8304496917338E-17 &  0.8300227385501E-17 &  0.8287393547857E-17 \\
 0.0666125 & 0.7647871188006E-17 & 0.7647866999040E-17 &  0.7639810367876E-17 &  0.7635810408153E-17 &  0.7623787342894E-17 \\
 0.0703125 & 0.7036141501167E-17 & 0.7036137510947E-17 &  0.7028596343733E-17 &  0.7024852160681E-17 &  0.7013598586415E-17 \\
 0.0741125 & 0.6473983501310E-17 & 0.6473979894812E-17 &  0.6466926621475E-17 &  0.6463424432608E-17 &  0.6452899021906E-17 \\
 0.0780125 & 0.5957648673312E-17 & 0.5957645571010E-17 &  0.5951052902029E-17 &  0.5947779099168E-17 &  0.5937940999571E-17 \\
               \hline
 0.0820125 & 0.5483587974654E-17 & 0.5483585440313E-17 &  0.5477426643500E-17 &  0.5474367918339E-17 &  0.5465177153440E-17 \\
 0.0861125 & 0.5048462721451E-17 & 0.5048460771207E-17 &  0.5042709880386E-17 &  0.5039853319997E-17 &  0.5031271090582E-17 \\
 0.0903125 & 0.4649148629125E-17 & 0.4649147241080E-17 &  0.4643779208428E-17 &  0.4641112365692E-17 &  0.4633101268488E-17 \\
 0.0946125 & 0.4282734755358E-17 & 0.4282733879064E-17 &  0.4277724681340E-17 &  0.4275235624615E-17 &  0.4267759803810E-17 \\
 0.0990125 & 0.3946518733255E-17 & 0.3946518298616E-17 &  0.3941845006054E-17 &  0.3939522351764E-17 &  0.3932547596255E-17 \\
 0.1035125 & 0.3637999386976E-17 & 0.3637999312224E-17 &  0.3633640127535E-17 &  0.3631473058861E-17 &  0.3624966856151E-17 \\
 0.1081125 & 0.3354867581199E-17 & 0.3354867779736E-17 &  0.3350802053473E-17 &  0.3348780327823E-17 &  0.3342711883700E-17 \\
 0.1128125 & 0.3094995961090E-17 & 0.3094996347097E-17 &  0.3091204574588E-17 &  0.3089318522133E-17 &  0.3083658753701E-17 \\
 0.1176125 & 0.2856428083343E-17 & 0.2856428576147E-17 &  0.2852892380277E-17 &  0.2851132895481E-17 &  0.2845854403773E-17 \\
 0.1225125 & 0.2637367314499E-17 & 0.2637367841750E-17 &  0.2634069945105E-17 &  0.2632428472874E-17 &  0.2627505500110E-17 \\
              \hline 
 0.1275125 & 0.2436165774447E-17 & 0.2436166274218E-17 &  0.2433090463444E-17 &  0.2431558981225E-17 &  0.2426967356852E-17 \\
 0.1326125 & 0.2251313525848E-17 & 0.2251313947789E-17 &  0.2248445032448E-17 &  0.2247016029429E-17 &  0.2242733108233E-17 \\
 0.1378125 & 0.2081428150094E-17 & 0.2081428455791E-17 &  0.2078752223555E-17 &  0.2077418677866E-17 &  0.2073423272820E-17 \\
 0.1431125 & 0.1925244803977E-17 & 0.1925244966678E-17 &  0.1922748136311E-17 &  0.1921503490986E-17 &  0.1917775802922E-17 \\
 0.1485125 & 0.1781606815719E-17 & 0.1781606819574E-17 &  0.1779276992832E-17 &  0.1778115131145E-17 &  0.1774636676911E-17 \\
 0.1540125 & 0.1649456852113E-17 & 0.1649456691077E-17 &  0.1647282304415E-17 &  0.1646197525071E-17 &  0.1642951065436E-17 \\
 0.1596125 & 0.1527828668478E-17 & 0.1527828344977E-17 &  0.1525798621745E-17 &  0.1524785614298E-17 &  0.1521755082618E-17 \\
 0.1653125 & 0.1415839438357E-17 & 0.1415838961857E-17 &  0.1413943865477E-17 &  0.1412997686281E-17 &  0.1410168118652E-17 \\
 0.1711125 & 0.1312682649388E-17 & 0.1312682034922E-17 &  0.1310912223441E-17 &  0.1310028272342E-17 &  0.1307385739798E-17 \\
 0.1770125 & 0.1217621544363E-17 & 0.1217620811106E-17 &  0.1215967593385E-17 &  0.1215141591308E-17 &  0.1212673134477E-17 \\
    \hline\hline
  \end{tabular}}
  \end{center}
  \end{table}
\newpage
\begin{table}[tbp]
   \caption{Photodetachment cross-section (in $cm^{2}$) of the negatively charged hydrogen ions in their ground $1^1S-$states
            for kinetic energies of photo-electron $E_e$ which are far from the maximum of the cross-section. All notations are the 
            same as in the two previous Tables.}
     \begin{center}
     \scalebox{0.80}{%
     \begin{tabular}{| c | c | c | c | c | c |}
      \hline\hline
        & ${}^{\infty}$H$^{-}$ &  ${}^{\infty}$H$^{-}$ & ${}^{3}$H$^{-}$ & ${}^{2}$H$^{-}$ & ${}^{1}$H$^{-}$ \\
           \hline
  $E_e$  & $\sigma(E_e) (N$ = 300) & $\sigma(E_e) (N$ = 350) & $\sigma(E_e) (N$ = 350) & $\sigma(E_e) (N$ = 350) & $\sigma(E_e) (N$ = 350) \\
          \hline    
 0.1830125 & 0.1129983081672E-17 & 0.1129982251607E-17 &  0.1128437545329E-17 &  0.1127665512883E-17  & 0.1125359079348E-17 \\
 0.1891125 & 0.1049152386308E-17 & 0.1049151483018E-17 &  0.1047707774635E-17 &  0.1046986011787E-17  & 0.1044830396360E-17 \\
 0.1953125 & 0.9745676610310E-18 & 0.9745667086430E-18 &  0.9732170153305E-18 &  0.9725420821878E-18  & 0.9705268701701E-18 \\
 0.2016125 & 0.9057155267702E-18 & 0.9057145490617E-18 &  0.9044523827004E-18 &  0.9038210814197E-18  & 0.9019365949053E-18 \\
 0.2080125 & 0.8421267615897E-18 & 0.8421257812580E-18 &  0.8409451144388E-18 &  0.8403544722382E-18  & 0.8385917194393E-18 \\
 0.2145125 & 0.7833724083041E-18 & 0.7833714463962E-18 &  0.7822666804163E-18 &  0.7817139336736E-18  & 0.7800645612302E-18 \\
 0.2211125 & 0.7290602220173E-18 & 0.7290592975088E-18 &  0.7280252323105E-18 &  0.7275078117097E-18  & 0.7259640599065E-18 \\
 0.2278125 & 0.6788314302517E-18 & 0.6788305597618E-18 &  0.6778623657516E-18 &  0.6773778824334E-18  & 0.6759325432339E-18 \\
 0.2346125 & 0.6323577798990E-18 & 0.6323569775261E-18 &  0.6314501692030E-18 &  0.6309964018707E-18  & 0.6296427796583E-18 \\
 0.2415125 & 0.5893388468558E-18 & 0.5893381241154E-18 &  0.5884885354307E-18 &  0.5880634183601E-18  & 0.5867952932704E-18 \\
            \hline
 0.2485125 & 0.5494995858601E-18 & 0.5494989517092E-18 &  0.5487027130754E-18 &  0.5483043249749E-18  & 0.5471159186943E-18 \\
 0.2556125 & 0.5125880996827E-18 & 0.5125875606145E-18 &  0.5118410774791E-18 &  0.5114676311479E-18  & 0.5103535750516E-18 \\
 0.2628125 & 0.4783736084176E-18 & 0.4783731686084E-18 &  0.4776731015090E-18 &  0.4773229342409E-18  & 0.4762782397414E-18 \\
 0.2701125 & 0.4466446011363E-18 & 0.4466442626341E-18 &  0.4459875086486E-18 &  0.4456590733307E-18  & 0.4446791043100E-18 \\
 0.2775125 & 0.4172071536235E-18 & 0.4172069165616E-18 &  0.4165905920736E-18 &  0.4162824489309E-18  & 0.4153628961306E-18 \\
 0.2850125 & 0.3898833972676E-18 & 0.3898832600957E-18 &  0.3893046847916E-18 &  0.3890154937331E-18  & 0.3881523509801E-18 \\
 0.2926125 & 0.3645101254581E-18 & 0.3645100851790E-18 &  0.3639667672004E-18 &  0.3636952807213E-18  & 0.3628848228715E-18 \\
 0.3003125 & 0.3409375250224E-18 & 0.3409375774283E-18 &  0.3404271996037E-18 &  0.3401722562045E-18  & 0.3394110186911E-18 \\
 0.3081125 & 0.3190280213342E-18 & 0.3190281612350E-18 &  0.3185485683200E-18 &  0.3183090864030E-18  & 0.3175938462811E-18 \\
 0.3160125 & 0.2986552267348E-18 & 0.2986554481751E-18 &  0.2982046350311E-18 &  0.2979796072405E-18  & 0.2973073656244E-18 \\
            \hline
 0.3240125 & 0.2797029828421E-18 & 0.2797032793025E-18 &  0.2792793799499E-18 &  0.2790678679205E-18  & 0.2784358337181E-18 \\
 0.3321125 & 0.2620644881721E-18 & 0.2620648527520E-18 &  0.2616661302361E-18 &  0.2614672597225E-18  & 0.2608728345722E-18 \\
 0.3403125 & 0.2456415032806E-18 & 0.2456419288630E-18 &  0.2452667658745E-18 &  0.2450797222305E-18  & 0.2445204865560E-18 \\
 0.3486125 & 0.2303436263476E-18 & 0.2303441057424E-18 &  0.2299909959384E-18 &  0.2298150199219E-18  & 0.2292887200228E-18 \\
 0.3570125 & 0.2160876327733E-18 & 0.2160881588421E-18 &  0.2157556988156E-18 &  0.2155900826944E-18  & 0.2150946187971E-18 \\
    \hline\hline
  \end{tabular}}
  \end{center}
  \end{table}

\begin{thebibliography}{10}
\bibitem{Cass} A well known Be-star is the $\gamma$ Cassiopeia which can easily be located and observed at the clear night sky in Northern Hemisphere.
\end{thebibliography}
\end{document}